\documentclass[twocolumn]{IEEEtran}
\usepackage{amsfonts}
\usepackage[T1]{fontenc}
\usepackage{color}
\usepackage{float}
\usepackage{mathrsfs}
\usepackage{amsmath,amssymb}
\usepackage{cases}
\usepackage{graphicx}
\usepackage[ruled,noend]{algorithm2e}
\usepackage{braket}
\usepackage{subfigure}
\usepackage{caption}

\makeatletter
\ifCLASSOPTIONcompsoc
\else
\fi
\@ifundefined{showcaptionsetup}{}{ \PassOptionsToPackage{caption=false}{subfig}}
\makeatother
\SetKwInput{KwCap}{Algorithm}
\SetKwInput{KwMD}{Method}
\SetAlgoCaptionSeparator{.}
\newtheorem{theorem}{Theorem}
\newtheorem{lemma}[theorem]{Lemma}

\newtheorem{definition}{Definition}
\newcommand{\cd}[1]{\ensuremath{\left|#1\right|}}

\begin{document}

\title{Multi-View 3D Video Multicast for Broadband IP Networks}

\author{\authorblockN{Ting-Yu Ho\authorrefmark{1}, Yi-Nung Yeh\authorrefmark{2}, and De-Nian Yang\authorrefmark{2}\authorrefmark{3}}\\
\authorblockA{\authorrefmark{1}Department of Industrial and Systems Engineering, University of Washington, Washington, USA\\
\authorrefmark{2}Institute of Information Science, Academia Sinica, Taipei, Taiwan\\
\authorrefmark{3}Research Center for Information Technology Innovation, Academia Sinica, Taipei, Taiwan}}
\maketitle

\begin{abstract}
With the recent emergence of 3D-supported TVs, video service providers now face an opportunity to provide high resolution multi-view 3D videos over IP networks. One simple way to support efficient communications between a video server and multiple clients is to deliver each desired view in a multicast stream. Nevertheless, it is expected that significantly increased bandwidth will be required to support the transmission of all views in multi-view 3D videos. However, the recent emergence of a new video synthesis technique called Depth-Image-Based Rendering (DIBR) suggests that multi-view 3D video does not necessarily require the transmission of all views. Therefore, we formulate a new problem, named Multi-view and Multicast Delivery Selection Problem (MMDS), and design an algorithm, called MMDEA, to find the optimal solution. Simulation results manifest that using DIBR can effectively reduce bandwidth consumption by $35\%$ compared to the original multicast delivery scheme.
\end{abstract}


\section{Introduction}

\label{sec:intro}

\IEEEPARstart{T}{elevision} with 4K and 3D-support were heralded as the future of television at the 2104 Consumer Electronics Show (CES), and many television manufacturers including Samsung, Sony, LG, and
Philips have introduced 3D Smart LED TV to markets. Internet video providers, such as YouTube and Netflix, now provide 3D videos and 3D live streaming service to users for Internet-ready 3DTVs. In contrast to traditional 3D
videos which offer the users only a single viewpoint, multi-view 3D videos allow the users to choose from a range of viewing angles. Currently the Digital Video Broadcasting (DVB) 3DTV standard supports multi-view 3D videos. In addition to DVB, a more flexible way to distribute 3D media is to stream over the Internet \cite{3D-Tekalp2007,3D-Gurler2011}. Several companies and
research teams have built demonstration systems for multi-view 3D
video service over Internet Protocol (IP) networks \cite%
{3D-Aksay2007,3D-Matusik2004}. Moreover, research and applications for 3D video broadcast and IP streaming services have been presented \cite{3D-Hewage2009,3D-Chung2010}, allowing IPTV Service companies to provide multi-view 3D video streaming over IP networks \cite{3D-Kim2008}. The mist straightforward way to support efficient communications between a
video server and multiple terminal users is to deliver every
view of a multi-view 3D video in a multicast stream. Nevertheless, while different users enjoy their preferred views, it is expected that the bandwidth requirements in the network will significantly increase to support
all views in multi-view 3D videos \cite{3D-Minoli2012,IPmulticast-Lou2007}.

Depth-Image-Based Rendering (DIBR) \cite{3D-Fehn2004} is one promising way
to remedy the bandwidth issue in the multi-view 3D video delivery. Because
adjacent views usually share many similar contents, the desired view of a
client can be synthesized from one nearby left view and one nearby right
view, and researchers in image processing and video coding have developed
sophisticated DIBR algorithms to ensure good synthesis quality by optimizing
the bit allocation between the texture and depth map among views \cite%
{3D-Mori2009,3D-Ndjiki-Nya2011}. Therefore, with the capability to render
arbitrary views, DIBR has been recognized as an efficient way to provide
Free Viewpoint Videos (FVV) applications \cite{3D-Smolic2011}, where each
client can arbitrarily specify the desired view. Equipped with DIBR in
clients, the bandwidth consumption in a network can be effectively reduced.

However, this approach is subject to several challenges. 1) To avoid the generation of unacceptable disoccluded areas in synthesized virtual views, the left and right views used to synthesize the desired view must be reasonably close to one another \cite{3D-Mori2009}. Different users desire different different views, and satisfying these demands require carefully selecting views for transmission so that the desired view of each user can be synthesized
with good quality. In other words, the quality constraint in DIBR specifies
that the left and right views are allowed to be at most $D$ views away
(i.e., $D-1$ views between them), to guarantee good quality of every
synthesized view between them. 2) To support more multi-view videos in IP networks, a simple approach is to minimize the bandwidth consumption by transmitting only the minimal number of views required. Nevertheless, since the current IP multicast routing protocols, PIM-SM \cite{3D-RFC, 3D-Wang2007}, exploit a
shortest-path tree for point-to-multipoint group communications, the network
bandwidth to deliver each view varies since each user may prefer a different
view. Moreover, to synthesize a view using DIBR, the user must receive two views instead of one, thus a more promising approach is to acquire the close left and right  views from nearby users in the corresponding
two multicast trees. However, selecting views for delivery to nearby presents a challenge and different view selections for various users results in different tree routing. Therefore, it is desired to have a smart view selection strategy to minimize the total bandwidth consumption in all multicast trees to provide scalable multi-view 3D video services over a network.

\begin{figure}[t]
\centering
\includegraphics[width=.75\linewidth]{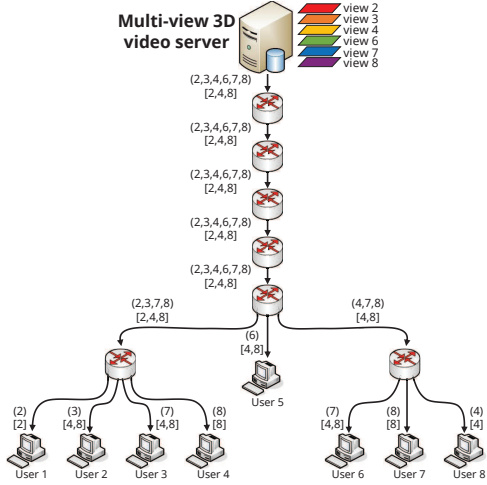}
\caption{Multi-view 3D video multicast routing.}
\label{fig:MMDS}
\end{figure}

Fig.~\ref{fig:MMDS} presents an illustrative example for efficient delivery
of a multi-view 3D video, which includes one video server, five routers and
eight client users. In this example, users 1 to 8 request the preferred
views 2,3,7,8,6,7,8, and 4, respectively. One intuitive way, called \emph{%
original multicast delivery scheme}, is to multicast each desired view to
each client directly, and the views transmitted in each link listed in the
parenthesis. The total bandwidth consumption is 45, where the \emph{total
bandwidth consumption} is the sum of the number of views delivered in every
edge (see Definition~\ref{def:cost}). In contrast, a more efficient way is
to exploit DIBR to reduce the bandwidth consumption. Take $D=4$ for an
example with the views transmitted in each link listed in the bracket. The
total bandwidth consumption can be effectively reduced to 32 by the
following selections: $2\mapsto (2,2)$, $3\mapsto (2,4)$, $4\mapsto (4,4)$, $%
6\mapsto (4,8)$, $7\mapsto (4,8)$, $8\mapsto (8,8)$, where $b\mapsto (a,c) $
represents that view $b$ is synthesized by views $a$ and $c$ if $a\neq c$;
otherwise view $b$ in $b\mapsto (b,b)$ is processed directly. With DIBR, it
is only necessary to deliver views 2,4, and 8 for all clients.

Based on the above observations, we make the first attempt to propose an
efficient view selection strategy for multi-view video delivery in IP
networks. We formulate a new optimization problem, called \emph{Multi-view
and Multicast Delivery Selection Problem} (MMDS), to minimize the total
bandwidth consumption for efficient multi-view 3D video multicast in IP
networks. We design an algorithm, called \emph{Multi-view and Multicast
Delivery Exploration Algorithm} (MMDEA), to find an optimal solution of the
MMDS problem. %In addition, we also propose a heuristic algorithm for the MMDS problem to support the dynamic switching of the desired views. Due to the space constraint, the heuristic algorithm can be found in \cite{CORR}.
Our simulation results manifest that with exploiting DIBR, the bandwidth
consumption can be effectively reduced by $35\%$, comparing to the original
multicast delivery scheme.
%In this paper, we assume that each user preferring the same view receive
%the same reference views from the server whether the preferred view
%is satisfied by DIBR or direct preferred view serving.
Note that layer encoding multicasting also enables the delivery of
multimedia contents to client communities in a cost-efficient manner and can
automatically adjust the transmission of the base layer and successive
layers according to the available bandwidth. However, the multi-view
transmission with DIBR needs to select the transmission views by examining
the preferred view of all clients as well as the topology of SPT, resulting
a more challenging issue.

The rest of the paper is organized as follows. Section~\ref%
{sec:problem-formulation} describes the system model and formulates the MMDS
problem. Section~\ref{sec:MMDEA} demonstrates the idea of MMDEA by first
considering two fundamental special cases and then extend it to the general
case.

%Section IV addresses a quick transmission update approach for dynamic view delivery.
Section~\ref{sec:simulation} presents the simulation results and we conclude
this paper in Section~V.

The rest of the paper is organized as follows. Section II describes the system model and formulates the MMDS problem. Section III demonstrates the idea of MMDEA by ﬁrst considering two fundamental special cases and then extend it to the general case. Section IV considers a generalization of the MMDS problem. Section V proposes a heuristic algorithm to support the quick switching of the desired views transmission. Section VI presents the simulation results and we conclude this paper in Section VII.

\section{Problem Formulation}

\label{sec:problem-formulation}

The network consists of a shortest path directed tree $T=(V,A,s)$ spanning a
video server and all clients, where $V$ and $A$ denote the set of nodes and
directed edges, respectively, and $s$ is the root of $T$, which is
considered to be the multi-view video server in the network. The set of
terminal nodes of $T$ is denoted by $\Omega _{T}$, which represents the set
of clients in the network. The directed path from $s$ to $t\in \Omega _{T}$
is denoted by $P_{s,t}$. Let $\mathcal{V}\footnote{%
For convenience, we assume that the views provided by the video server are
finite, distinct, and are presented by consecutive positive integers.}%
\subseteq \mathbb{N}$ denote the universal set of views in a multi-view 3D
video, and $\rho _{T}\colon \Omega _{T}\rightarrow \mathcal{V}$ denotes a
\emph{preferred-view function}, i.e., each terminal node $t$ selects a
desired view $\rho _{T}(t)$ from $\mathcal{V}$. Let $\mathcal{V}_{\rho
}\subseteq \mathcal{V}$ denote the set of all desired views by all clients.

Let $D$ denote the DIBR quality constraint\footnote{%
The DIBR quality constraint $D$ is a positive integer with $D\geq 2$.}. The
MMDS problem aims to find an optimal view-selection function (i.e., assigns
a view or two nearby views to each client) to minimize the total bandwidth
consumption in the network. More specifically, given the set of preferred
views $\mathcal{V}_{\rho }$, let $\theta \colon \mathcal{V}_{\rho
}\rightarrow \mathcal{V}\times \mathcal{V}$ be a \emph{view-selection
function} that assigns each preferred view $v$ in $\mathcal{V}_{\rho }$ an
ordered pair of views $(\theta (v).\ell ,\theta (v).r)$ from $\mathcal{V}$,
where $\theta (v).\ell =\theta (v).r$ or $\theta (v).\ell <\theta (v).r$.
For a view-selection function $\theta $, we say that \emph{$\theta $
satisfies $\mathcal{V}_{\rho }$ with respect to $D$} if $\theta $ satisfies
the following three conditions: 1) $\theta $ fits the DIBR quality
constraint, i.e., $0\leq \theta (v).r-\theta (v).\ell \leq D$ for all $v\in
\mathcal{V}_{\rho }$; 2) the left and right views $\theta (v).\ell $ and $%
\theta (v).r$ (i.e., $\theta (v).r\neq \theta (v).\ell $) cannot be further
synthesized by other views. Specifically, if $\theta (v).r>\theta (v).\ell $%
, $\theta (v^{\prime }).\ell =\theta (v^{\prime }).r$ must hold for $%
v^{\prime }=\theta (v).\ell $ or $v^{\prime }=\theta (v).r$. 3) $\theta $
has no crossing view selections,
%\footnote{%
%The variation of crossing view selections will be studied in \cite{CORR}.}
i.e., if $\theta (v).r\neq \theta (v).\ell $ for some view $v$, no view $%
v^\prime$ can be assigned $(\theta (v^\prime).\ell,\theta (v^\prime).r)$
with $\theta (v).\ell <\theta (v^\prime).\ell<\theta (v).r$ or $\theta
(v).\ell <\theta (v^\prime).r<\theta (v).r$. We formulate the MMDS problem
as follows.

\begin{definition}
\label{def:cost} Given a rooted tree $T=(V,A,s)$, a universal view set $%
\mathcal{V}$, a preferred-view function $\rho _{T}$ and thus $\mathcal{V}%
_{\rho }$, and the DIBR quality constraint $D$, the MMDS problem is to find
a view-selection function $\theta $ such that $\theta $ satisfies $\mathcal{V%
}_{\rho }$ with respect to $D$, and the \emph{total bandwidth consumption}
defined in (\ref{eq:bandwidth}) is minimized.
\begin{equation}
cost(\theta )=\sum_{e\in A}\Bigm|\!\!\mathop{\bigcup_{t\in\Omega_T}}_{e\in
P_{s,t}}\!\!\set{\theta(\rho_T(t)).\ell,\theta(\rho_T(t)).r}\!\Bigm|.
\label{eq:bandwidth}
\end{equation}
\end{definition}

The cost in \eqref{eq:bandwidth} indicates that every view selected for the
clients will be counted once on every edge of the paths from the root to the
clients. Therefore, the objective function encourages two or more clients
that share many common edges in their paths from the root to exploit the
same views, while each view can be directly processed by a client or be
regarded a the left or right view for synthesis with DIBR. Let $\theta
^{\ast }$ denote an optimal view-selection function to the MMDS problem.
After $\theta ^{\ast }$ is decided, the set of views required to be
transmitted at the video server $s$ will be
\begin{equation}
\mathcal{V}^{\ast }=\bigcup_{v\in \mathcal{V}_{\rho }}\set{\theta^*(v).\ell,%
\theta^*(v).r}.  \label{eq:view-at-video-server}
\end{equation}

In this paper, we explore the fundamental problem of
providing efﬁcient multi-view 3D multicasts over broadband IP
networks, where each client has sufﬁcient bandwidth to receive
two views. The problem with some clients only able to receive
one view is a special case of the problem, by enforcing that
the desired view cannot be synthesized.

\section{Algorithm Design}

\label{sec:algo}

An intuitive approach to address the MMDS problem is to iteratively select
the view that can serve the most number of clients in order to reduce the
total bandwidth consumptions. Nevertheless, the strategy does not carefully
examine the network structure and identify the closeby clients that share a
long common path from the root. In addition, it does not consider the
desired views of multiple clients jointly to find out the views that can be
shared by those client as the left and right views for synthesis with DIBR.
As $D$ and the number of views increase, the problem become more challenging
since it will impose much more choices during the selection of views for
each client. As a result, instead of trying all possible choices of views to
minimize the total bandwidth consumption, we present an algorithm called
\emph{Multi-view and Multicast Delivery Exploration Algorithm} (MMDEA) to
systematically derive an optimal solution for the MMDS problem with dynamic
programming. In the following, we will first present the algorithm with $D=2$
and 3 and then extend it to the general case. The algorithm can be
implemented by the SDN controller or the video server, where the routing
information of the shortest-path tree is able to be acquired by ICMP
traceroute.
%Due to space constraint, the support of the dynamic join and
%leave of clients, together with the change of preferred views, is presented
%in \cite{CORR}.

\subsection{Dynamic Programming Formulation}

To effectively minimize the total bandwidth consumption, we propose MMDEA
based on dynamic programming. MMDEA first divides the desired views set $%
\mathcal{V}_{\rho }$\footnote{%
To avoid ambiguity, we use $v_{i}$ to represent view $i$ in the rest of this
paper. Moreover, assume that the views in $\mathcal{V}_{\rho }$ are listed
in the non-decreasing order.} into multiple non-overlapping maximal segments
$\mathcal{V}_{\rho }^{1},\ldots ,\mathcal{V}_{\rho }^{n}$ such that the
\emph{gap} (the largest value of $\left\vert v_{i}-v_{j}\right\vert $ with
no view from $v_{i}$ to $v_{j}$ in $\mathcal{V}_{\rho }$) in each segment is
no larger than $D$. For example, if $D=3$ and $\mathcal{V}_{\rho }=%
\set{1,2,3,5,9,10,15,17,18}$, then we can divide $\mathcal{V}_{\rho }$ into
three segments: $\mathcal{V}_{\rho }^{1}=\set{1,2,3,5}$, $\mathcal{V}_{\rho
}^{2}=\set{9,10}$ and $\mathcal{V}_{\rho }^{3}=\set{15,17,18}$.

For $m\leq k$, let $c_{m,k}$ denote the minimum cost of a view-selection
function $\theta _{m,k}^{\ast }$ with the set of desired views as $%
\set{v_m,v_{m+1},\ldots,v_k}\cap \mathcal{V}_{\rho }$, where the two
boundary views $v_{m}$ and $v_{m}$ must be selected in $\theta _{m,k}^{\ast
} $. In other words, $c_{m,k}$ is the minimum total bandwidth consumption to
serve the clients with the desired views from $v_{m}$ to $v_{k}$, and $v_{m}$
and $v_{k}$ are the boundary views and thus need to be transmitted directly
or be generated by views using DIBR synthesis. The cost induced from any
views not in $\set{v_m,v_{m+1},\ldots,v_k}\cap \mathcal{V}_{\rho }$ is not
included in $c_{m,k}$. Consequently, the minimum total bandwidth consumption
to the MMDS problem is
\begin{equation}
\sum_{i=1}^{n}c_{m_{i},M_{i}}  \notag
\end{equation}%
for $\mathcal{V}_{\rho }^{i}=\set{v_{m_i},\ldots,v_{M_i}}$, where $v_{m_{i}}$
and $v_{M_{i}}$ denote the minimum and the maximum view in $\mathcal{V}%
_{\rho }^{i}$, respectively.

It is worth noting that, although only the views in $\mathcal{V}_{\rho }$
are desired, some views in $\mathcal{V}\setminus \mathcal{V}_{\rho }$ may
still be selected in the solution for synthesis with DIBR in order to
minimize the total bandwidth consumption. For simplicity, we will focus on
deriving $c_{m,M}$ for each segment $\mathcal{V}_{\rho }^{i}=%
\set{v_m,\ldots,v_M}$ in the rest of this paper. In the following, we first
explore the fundamental cases with $D=2$ and 3 to derive $c_{m,k}$
systematically for each $k\in \set{m,{m+1},\ldots,M}$.

\subsection{Special Case}

In this section, we aim at establishing the recursive relation of $c_{m,k}$
for DIBR with $D=2$ and 3. We first consider the case of $D=2$. Two
fundamental costs are involved to find $c_{m,k}$. The first one is $c_{k,k}$%
, which represents the total bandwidth consumption to multicast view $v_{k}$
to every client that subscribes the view. In other words, $c_{k,k}$ is the
cost of the multicast tree to span all clients that subscribe $v_{k}$. In
addition, for any subset $\mathcal{V}^{\prime }$ of $\mathcal{V}_{\rho }$
and two boundary views $v_{\ell }$ and $v_{r}$ such that $v_{r}-v_{\ell
}\leq D$ and $v_{\ell }<v<v_{r}$ for every view $v\in \mathcal{V}^{\prime }$%
, let $\Phi _{(v_{\ell },v_{r})}^{\mathcal{V}^{\prime }}$ denote the \emph{%
expansion-cost function}, which is additional bandwidth consumption to
multicast view $v_{\ell }$ and $v_{r}$ to every client that subscribes $v\in
\mathcal{V}^{\prime }$ between $v_{\ell }$ and $v_{r}$, in order to
synthesize view $v$ with DIBR, if the mutlicast tree for the views in $%
\{v_{m},\ldots ,v_{l},v_{r}\}$ has been constructed. In other words, $\Phi
_{(v_{\ell },v_{r})}^{\mathcal{V}^{\prime }}$ is the additional cost
required to expand the multicast tree that has spaned other clients
subscribing views in $\{v_{m},\ldots ,v_{l},v_{r}\}$ to reach the clients
subscribing the views in $\mathcal{V}^{\prime }$. For
simplicity, let $\Phi _{(v_{\ell },v_{r})}^{\mathcal{V}^{\prime }}=0$ if $%
\mathcal{V}^{\prime }\cap \mathcal{V}_{\rho }=\emptyset $. In the following,
we first define $c_{k}$ as follows.
\begin{numcases}{c_k\!=\!\!}
c_{k,k} \hspace{-0.5cm}& if $v_k\in\mathcal{V}_\rho$\notag\\
\infty  \hspace{-0.5cm}& if $v_k\not\in\mathcal{V}_\rho$ and $v_k$ is not generated by any view\notag\\
0       \hspace{-0.5cm}& if $v_k\not\in\mathcal{V}_\rho$ and $v_k$ is generated by some views.\notag
\end{numcases}
%define $c_{k,k}=\infty$ if $v_k$ cannot be generated with respect to $D$ from the synthesized views
%by any view in $\mathcal{V}_\rho^{m,k}$, otherwise, define $c_{k,k}=0$.
%In other words, $c_{k,k}=\infty$ if all views
%from $v_{k-D+1}$ to $v_{k-1}$ are missing from $\mathcal{V}_\rho$.

$\mathbf{D=2.}$ Let $c_{m,k}^{0}$ denote the bandwidth consumption to serve
the clients with the desired views from $v_{m}$ to $v_{k}$, where $v_{k}$ is
employed to serve the clients subscribing $v_{k}$ only. By contrast, let $%
c_{m,k}^{1}$ denote the bandwidth consumption for the same clinets, but $%
v_{k}$ here is also exploited to serve the clients for synthesizing $v_{k-1}$
with DIBR. The following lemma shows that $c_{m,k}$ can be obtained by
comparing $c_{m,k}^{0}$ and $c_{m,k}^{1}$, where the proof explains the
detailed multicast opeations for all possible cases.

\begin{lemma}
\label{lem:D2} For $D=2$ and $k\in\set{m,m+1,\ldots,M}$, let $J=\set{0,1}$,
and we have %if $v_k\not\in\mathcal{V}_\rho$, then
%$c_{m,k} = c_{m,k-2}+\Phi_{(v_{k-2},v_k)}^{\{v_{k-1}\}}$, otherwise
\begin{numcases}{c_{m,k}\!=\!\min\!}
\!\!c^0_{m,k}\!=\!\min\{c_{m,k-1},c_{m,k-2}\}+c_k\label{eq:DIBRTWOA}\\
%\min_{\theta\in\overline{\theta}_{m,k-2}}\{cost(\theta)+c_k+\Phi_{(v_{k-2},v_k)}^{\{v_{k-1}\}}\}.\quad\label{eq:DIBRTWOB}\\
\!\!c^1_{m,k}\!=\!\min_{j\in J}\{c_{m,k-2}^j+c_k+\Phi_{(v_{k-2},v_k)}^{\{v_{k-1}\}}\}.\quad\label{eq:DIBRTWOB}
\end{numcases}
\end{lemma}

\begin{IEEEproof}
We prove the lemma by induction on $k$. The result holds clearly for $k=m$.
Suppose it holds $c_{m,k'}$ for every $k'<k$.
Assume that $v_{k-1}\in\mathcal{V}_\rho$.
There are two possible cases as follows.

\emph{Case 1}: view $v_{k}$ is not involved in the view synthesis.
This implies that no view from $v_m$ to $v_{k-1}$ is synthesized by $v_k$.
If $v_{k-1}\in\mathcal{V}_\rho$, then we have $v_{k-1}\mapsto (v_{k-1},v_{k-1})$,
implying that $c^0_{m,k}=c_{m,k-1}+c_k$.
Alternatively, for $v_{k-1}\not\in\mathcal{V}_\rho$,
since the gap of $\mathcal{V}_\rho$ is no larger than $D$,
$v_{k-2}\in \mathcal{V}_\rho$ and $v_k\in\mathcal{V}_\rho $ hold, and thus we have $v_{k}\mapsto (v_{k},v_{k})$.
On the other hand, there are two possible cases for $v_{k-2}$, i.e., $v_{k-2}\mapsto (v_{k-2},v_{k-2})$ or $v_{k-2}\mapsto (v_{k-3},v_{k-1})$.
In the former case, $c^0_{m,k}=c_{m,k-2}+c_k$ holds;
in the latter case, $c^0_{m,k}=c_{m,k-1}+c_k$ holds.

\emph{Case 2}: view $v_{k}$ is involved in the synthesis for $v_{k-1}$.
In this case, we have $v_{k-1}\in\mathcal{V}_\rho$ and $v_{k-1}\mapsto (v_{k-2},v_k)$.
Note that views $v_{k-2}$ and $v_k$ cannot be further synthesized
by other views and thus need be transmitted directly if they are in $\mathcal{V}_\rho$.
If $v_{k-2}$ is not exploited in the view synthesis,
we have $c^1_{m,k}=c_{m,k-2}^0+c_k+\Phi_{(v_{k-2},v_k)}^{\{v_{k-1}\}}$;
otherwise, $c^1_{m,k}=c_{m,k-2}^1+c_k+\Phi_{(v_{k-2},v_k)}^{\{v_{k-1}\}}$,
implying that \eqref{eq:DIBRTWOB} holds.
Since $c_{m,k}$ is a minimization, the smaller one of the above two cases
is the minimum cost of $c_{m,k}$. The lemma follows.
\end{IEEEproof}

After finding the minimum cost $c_{m,k}$ with the above recursive relation,
the optimal view-selection function $\theta _{m,k}^{\ast }$ can be obtained
from $c_{m,k}$ by backtracking with \eqref{eq:DIBRTWOA} and %
\eqref{eq:DIBRTWOB} as follows.

\emph{Case 1:~$c_{m,k}$ is derived from $c_{m,k}^{0}$ in \eqref{eq:DIBRTWOA}.%
} If $c_{m,k}^{0}=c_{m,k-1}+c_{k}$, we set $v_{k}\mapsto (v_{k},v_{k})$,
i.e., $v_{k}$ is transmitted directly. If $v_{k-1}\in \mathcal{V}_{\rho }$,
we set $v_{k-1}\mapsto (v_{k-1},v_{k-1})$, i.e., $v_{k-1}$ is also
transmitted directly. Afterwards, $c_{m,k-1}$ is processed similarly to find
$\theta _{m,k-1}^{\ast }$. On the other hand, if $v_{k-1}\not\in \mathcal{V}%
_{\rho }$, we set $v_{k-2}\mapsto (v_{k-3},v_{k-1})$ because it is more
bandwidth efficient to multicast view $v_{k-1}$ for $v_{k-2}$, instead of
directly transmitting $v_{k-2}$. Afterwards, $c_{m,k-2}$ is processed
similarly to find $\theta _{m,k-2}^{\ast }$. By contrast, if $%
c_{m,k}^{0}=c_{m,k-2}+c_{k}$, $v_{k}\in \mathcal{V}_{\rho }$ and $%
v_{k-1}\not\in \mathcal{V}_{\rho }$ must hold, and we have $v_{k-2}\mapsto
(v_{k-2},v_{k-2})$ and $v_{k}\mapsto (v_{k},v_{k})$, respectively, i.e.,
views $v_{k-2}$ and $v_{k}$ are transmitted directly. Afterwards, $c_{m,k-2}$
is processed similarly to find $\theta _{m,k-2}^{\ast }$.

\emph{Case 2:~$c_{m,k}$ is derived from $c_{m,k}^{1}$ in \eqref{eq:DIBRTWOB}.%
} Suppose $c_{m,k}^{1}=c_{m,k-2}^{j}+c_{k}+\Phi
_{(v_{k-2},v_{k})}^{\{v_{k-1}\}}$ for some $j\in \set{0,1}$. We set $%
v_{k-1}\mapsto (v_{k-2},v_{k})$ and $v_{k}\mapsto (v_{k},v_{k})$ for $%
v_{k-1},v_{k}\in \mathcal{V}_{\rho }$. In other words, $v_{k-1}$ is
synthesized from the two neighbor views. Afterwards, $c_{m,k-2}^{j}$ is
processed similarly to find $\theta _{m,k-2}^{\ast }$.

$\mathbf{D=3.}$ For $v_{k}$, only $v_{k-2}$ and $v_{k-1}$ can exploit $v_{k}$
for synthesis with DIBR. The possible cases for $v_{k-2}$ include $%
v_{k-2}\mapsto (v_{k-2},v_{k-2})$ (non-synthesis), $(v_{k-3},v_{k-1})$, or $%
(v_{k-3},v_{k})$, while $v_{k-1}\mapsto (v_{k-1},v_{k-1})$ (non-synthesis), $%
(v_{k-2},v_{k})$, or $(v_{k-3},v_{k})$ are also possible. Although there are
nine combinations to jointly examine $v_{k-2}$ and $v_{k-1}$, it is
necessary to examine only three of them. The first reason is that a selected
view cannot be further synthesized. For example, for $v_{k-2}\mapsto
(v_{k-3},v_{k-1})$, view $v_{k-3}$ and $v_{k-1}$ cannot be further
synthesized. Secondly, no cross synthesis is allowed. For example,~$%
v_{k-2}\mapsto (v_{k-2},v_{k-2})$ and $v_{k-1}\mapsto (v_{k-3},v_{k})$ are
not allowed to o-cexist simultaneously since the view synthesis of view $%
v_{k-1}$ cross $v_{k-1}$, which is transmitted directly. Thirdly, the combinations that do
not exploit $v_{k}$ for synthesis with DIBR has been considered when we
derive $c_{m,k-1}$, such as $v_{k-2}\mapsto (v_{k-2},v_{k-2})$ and $%
v_{k-1}\mapsto (v_{k-1},v_{k-1})$.

Specifically, Table \ref{tab:DIBR3} summarizes the new notations for $D=3$.
Let $c_{m,k}^{1}$ denote the bandwidth consumption to serve the clients
with the desired views from $v_{m}$ to $v_{k}$, where $v_{k}$ is employed to
synthesize $v_{k-1}$. Let $c_{m,k}^{2}$ denote the bandwidth consumption
for the same clinets, but $v_{k}$ here is exploited to synthesize both $%
v_{k-1}$ and $v_{k-2}$. \renewcommand{\arraystretch}{1.5}
\begin{table}[t]
\caption{The synthesis combinations in the computation of $c_{m,k}$ for $D=3$%
.}
\label{tab:DIBR3}\centering
\begin{tabular}{ccc}
\hline
& $v_{k-2}$ & $v_{k-1}$ \\ \hline
$c^{1}_{m,k}$ & non-synthesis & $(v_{k-2},v_k)$ \\
$c^{2}_{m,k}$ & $(v_{k-3},v_k)$ & $(v_{k-3},v_k)$ \\ \hline
\end{tabular}%
\end{table}
\renewcommand{\arraystretch}{1} Thus, $c_{m,k}$ for $D=3$ can be obtained by
the following recursive relation.
%Since it is extended by the proof of the previous lemma, the detailed proof is presented in \cite{CORR} due to the space constraint.

\begin{lemma}
For $D=3$, $k\in\set{m,m+1,\ldots,M}$, let $J=\{0,1,2\}$, and we have
\begin{numcases}{c_{m,k}\!=\!\min\!\!}
\!\!c^{0}_{m,k}\!=\!\min\{c_{m,k-1},c_{m,k-2},c_{m,k-3}\}+c_k \label{eq:DIBR3-0}\\
\!\!c^{1}_{m,k}\!=\!\min_{j\in J}\{c^j_{m,k-2}+c_k+\Phi_{(v_{k-2},v_k)}^{\{v_{k-1}\}}\}\label{eq:DIBR3-01}\\
\!\!c^{2}_{m,k}\!=\!\min_{j\in J}\{c^j_{m,k-3}+c_k+\Phi_{(v_{k-3},v_k)}^{\{v_{k-2},v_{k-1}\}}\}.\hspace{.8cm}\label{eq:DIBR3-12}
\end{numcases}
\end{lemma}

%Equation~\eqref{eq:DIBR3-01} corresponds to the following view selections:
%$v_{k-2}\mapsto (v_{k-2}, v_{k-2})$ and $v_{k-1}\mapsto (v_{k-2}, v_{k})$, i.e.,
%view $v_{k-2}$ is transmitted directly, while  view $v_{k-1}$ must be synthesized with DIBR.
%On the other hand, equation~\eqref{eq:DIBR3-12} corresponds to the following view selections
%$v_{k-2}\mapsto (v_{k-2}, v_{k})$ and $v_{k-1}\mapsto (v_{k-2}, v_{k})$, i.e.,
%both $v_{k-2}$ and $v_{k-1}$ are synthesized with DIBR by using $v_{k-3}$ and $v_{k}$.

\subsection{General Case}

\label{sec:MMDEA}

In last section, we have established the recursive formulas to derive $%
c_{m,k}$ for $k\in \set{m,m+1,\ldots,M}$ with $D=2$ and $3$. However, when $%
D $ grows, the number of combinations required to be examined grows rapidly.
The reason is that during the derivation of $c_{m,k}$, all views $%
v_{k-D+1},v_{k-D+2},\ldots ,v_{k-1}$ are able to select $v_{k}$ for
synthesis with DIBR. Therefore, it becomes much more difficult to derive $%
c_{m,k}$.
Algorithm 1 presents the pseudocode of MMEDA. The input
parameters include a computed single-source shortest path rooted tree $%
T=(V,A,s)$, a universal view set $\mathcal{V}$ provided by the video server,
a preferred-view function $\rho_T$ which assigns each terminal nodes of $T$
a desired view from $\mathcal{V}$, and the DIBR quality constraint $D$.
MMDEA determines the minimum total bandwidth consumption $cost(\theta^*)$
of a view-selection function $\theta^*$ such that $\theta^*$ satisfies
$\mathcal{V}_\rho$ with respect to $D$.
In the following, we present \emph{Multi-view and Multicast Delivery
Exploration Algorithm} (MMDEA), which includes two stages:
Initialization and Exploration.
The first stage initializes and identifies the service range for all
desired views by the clients. The second stage explores each segment of the
service range separately and consider each possible view selection combinations
to determine the minimum total bandwidth consumption in the network.

%-----------------------------------------------------------------------------
%
\DecMargin{1em}
\begin{algorithm}[!t]
\label{algo: MMDEA}
\SetNlSty{}{}{:}
\SetNlSkip{0.05em}
\SetEndCharOfAlgoLine{;}
%\scriptsize
\caption{Multi-view and Multicast Delivery Exploration Algorithm (\textsc{MMDEA})}
\KwIn{A rooted tree $T=(V,A,s)$, a universal view set $\mathcal{V}$,
a preferred-view function $\rho_T$, and the DIBR quality constraint $D$.}
\KwOut{The minimum total bandwidth consumption $cost(\theta^*)$ of a view-selection function $\theta^*$
which satisfies $\mathcal{V}_\rho$ with respect to $D$.}
\KwMD{}
%\scriptsize
\tcp*[h]{Initialization stage}\\
Identify the service range $\mathcal{V}_\rho\gets \mathcal{V}_\rho^1\cup\cdots\cup\mathcal{V}_\rho^n$\;
$cost(\theta^*)\gets 0$\;
\tcp*[h]{Exploration stage}\\
\ForEach{segment $\mathcal{V}_\rho^i \gets \{v_m,\ldots,v_M\}$}{
    \For{$k=m$ \KwTo $M$}{
        $c_{m,k}^0\gets \min\{c_{m,k-1},c_{m,k-2},\ldots,c_{m,k-D}\}+c_k$\;
        %\Indentp{-1.2em}
        $J\gets\{0,1,2,\ldots,\min\{D,k-m\}-1\}$\;
        \For{$d=2$ \KwTo $\min\{D,k-m\}$}{
            %\Indentp{-1em}
            %$I_d\gets \set{v_{\min\{m,k-D\}+1},\ldots,v_{\min\{m,k-d\}-1}}$\;
            $E_d\gets \set{v_{\min\{m,k-d\}+1},\ldots,v_{k-1}}$\;
%            \ForEach{$v\in I_d\cap \mathcal{V}_\rho$}{
%                %\Indentp{-1.2em}
%               % \scalebox{0.7}{
%               \tcp*[h]{\scalebox{0.75}{The possible selections for view $v\in I_d$}}\\
%                $\Theta(v)\gets \{\theta(v)=(v_\ell,v_r) \mid r-\ell\leq D,$\\
%                \hspace{1.2cm} $r\leq k,v_m\leq v_\ell\leq v\leq v_r,v_r\notin E_d\, \}$\;
%                %}
%            }
%            $\Gamma_d\gets$ \begin{minipage}[t]{6cm}{all possible view selection combinations by the views
%            in $I_d$ such that each selection combination satisfies $\mathcal{V}_\rho$ w.r.t. $D$;}\end{minipage}
            $c_{m,k}^d \gets \min_{j\in J}\Big\{c^j_{m,k-d} + c_k $\\
                 %\hspace{1.25cm}$+\sum_{v\in I_d,\theta(v)\in\Gamma}\Phi^{\{v\}}_{\{\theta(v).\ell,\theta(v).r\}}$\\
                 \hspace{1.25cm}$+\sum_{v\in E_d}\Phi^{\{v\}}_{\{v_{k-d},v_k\}}\Big\}$\;
            $\theta_{m,k}^d \gets \theta_{m,k-d}^j$\\
            \hspace{1.2cm}$\cup \set{v\mapsto (v_{k-d},v_k)\mid v\in E_d\cap\mathcal{V}_\rho}$\\
            \hspace{1.2cm}$\cup \set{v_k\mapsto (v_k,v_k)\mid v_k\in \mathcal{V}_\rho}$\;

        }
        $\overline{\theta}_{m,k}\gets \bigcup_{d=0}^{\min\{D,m-k\}-1} \theta_{m,k}^d$\;

        %\ForEach{$d$  {\bf and} $\Gamma\in \Gamma_d, $}{
%            \If{$F(d,\Gamma)\cap \{v_{k-D},\ldots,v_{k-d}\}=v_{k-d}$}{
%                $c_{m,k}^\Gamma \gets \min_{j\in J}\Big\{c^j_{m,k-d} + c_k $\\
%                 \hspace{1.25cm}$+\sum_{v\in I_d,\theta(v)\in\Gamma}\Phi^{\{v\}}_{\{\theta(v).\ell,\theta(v).r\}}$\\
%                 \hspace{1.25cm}$+\sum_{v\in E_d}\Phi^{\{v\}}_{\{v_{k-d},v_k\}}\Big\}$\;
%            }
%            \Else{
%                $c_{m,k}^\Gamma \gets \min_{F\in \mathcal{F}}\Big\{c^F_{m,k-d} + c_k $\\
%                 \hspace{1.25cm}$+\sum_{v\in I_d,\theta(v)\in\Gamma}\Phi^{\{v\}}_{\{\theta(v).\ell,\theta(v).r\}}$\\
%                 \hspace{1.25cm}$+\sum_{v\in E_d}\Phi^{\{v\}}_{\{v_{k-d},v_k\}}\Big\}$\;
%            }
%        }
        %$\theta_{m,k}^0\gets \theta_{m,k-1}\oplus \set{v_k\mapsto (v_k,v_k)\mid \hbox{if }v_k\in\mathcal{V}_\rho}$\;
        %$\overline{\theta}_{m,k}\gets \overline{\theta}_{m,k} \bigcup \theta_{m,k}^0$\;
        $c_{m,k}\gets \min\Big\{c_{m,k}^0,c_{m,k}^{d}\mid $\\
        \hspace{1.2cm} $d\in\set{1,2,\ldots,\min\{D,k-m\}-1}\Big\}$\;
        %$\theta_{m,k}\gets \arg\min c_{m,k}$\;
        %$\theta_{m,k}\gets \theta_{m,k}^d$ with $c_{m,k}^d=c_{m,k}$\;
        %$\overline{\theta}_{m,k}\gets \bigcup_d \overline{\theta}_{m,k}^d$\;
    }
    %$\theta_i^*\gets \theta_{m,M}$\;
    $cost(\theta^*)\gets cost(\theta^*)+c_{m,M}$\;
}
\Return $cost(\theta^*)$\;
\end{algorithm}

\subsubsection{Initialization Stage}

In the initialization stage, it is necessary to identify the service range
based on the preferred-view function to ensure the each subscribed view is
able to be directly transmitted or synthesized by other views. Therefore,
the same as the approach described for $D=2$, it can be achieved by first
sorting the desired views in non-decreasing order, and then by dividing the
desired views set $\mathcal{V}_{\rho }$ into multiple non-overlapping
maximal segments $\mathcal{V}_{\rho }^{1},\ldots ,\mathcal{V}_{\rho }^{n}$
such that the gap in each segment $\mathcal{V}_{\rho }^{i}$ is no larger
than $D$.
%If any two desired views, say, $v_a$ and $v_b$ with at most $D-1$ views between
%them are satisfied by transmitting $v_{a}$ and $v_{b}$ directly,
%all desired views from $v_a$ to $v_b$ can be served by $v_a$ and $v_b$ and
%are covered by the same view range thanks to DIBR.
%On the other hand, when $v_a$ or $v_b$ is synthesized by other two reference views,
%at least one view between $v_a$ and $v_b$ must be transmitted, and
%thereby other desired views between $v_a$ or $v_b$ or can be served.

\subsubsection{Exploration Stage}

Initialization stage defines the service range to satisfy the clients. In
this stage, each segment $\mathcal{V}_{\rho }^{i}$ is horizontally explored
separately in order to pursuit the minimum total bandwidth consumption in
the network. More specifically, the goal of this stage is to derive $c_{m,M}$
for each segment $\mathcal{V}_{\rho }^{i}=\set{v_m,\ldots,v_M}$, which
represents the minimum total bandwidth consumption to serve all clients that
subscribe views from $v_{m}$ to $v_{M}$. MMDEA explores $\mathcal{V}_{\rho
}^{i}$ systematically and derive $c_{m,k}$ for all $k\in \set{m,m+1,\ldots,M}
$ according to the derived values of $c_{m,k-D},c_{m,k-D+1},\ldots ,c_{k-1}$
and $c_{k}$. This is because when $v_{k}$ is involved in the computation of $%
c_{m,k}$, only the views $v_{m,k-D+1},v_{m,k-D+2},\ldots ,v_{k-1}$ can
select $v_{k}$ for synthesis with DIBR. In addition, the difficulty lies in
that the choices for the views from $v_{k-D+1}$ to $v_{k-1}$ may affect the
choices for the views from $v_{m}$ to $v_{k-D}$. To derive $c_{m,k}$
correctly, it is necessary to record all costs obtained in the computation
of $c_{m,k}$ for further examining in the wider service ranges in order to
minimize the total bandwidth consumption.

\begin{figure}[tb]
\centering
\includegraphics[width=.85\linewidth]{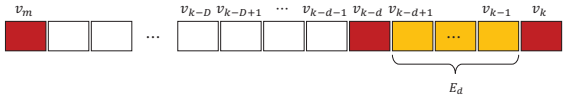}
\caption{An illustration of exploration stage. Those views in red color
(e.g., $v_m,v_{k-d},v_k$) are transmitted directly, while %The views in $I_d$
%(green color) may or may not be synthesized with DIBR, while
the views in $E_d$ (yellow color) are forced to select $(v_{k-d},v_k)$ for synthesis with
DIBR.}
\label{fig:exploreration-stage}
\end{figure}

The notion of exploration stage goes as follows. If $v_{k}$ is not exploited
to synthesize any other view, clearly $c_{m,k}=c_{m,k}^{0}=\min
\{c_{m,k-1},c_{m,k-2},\ldots ,c_{m,k-D}\}+c_{k}$, such as Eq. (5) for $D=3$.
Otherwise, it is necessary to examine different view selection combinations
that exploit $v_{k}$ for synthesis with DIBR. To find $c_{m,k}$ in this
case, MMDEA sequentially examines the case that a view $v_{k-d}$, is
transmitted, where $D\geq d\geq 2$. In addition, every other view between $%
v_{k-d}$ and $v_{k}$ is synthesized from the two views accordingly. For
example, when $D=3$, $v_{k-2}$ and $v_{k-3}$ are examined sequentially and
assumed to be transmitted, as explained in Eq. (6) and Eq. (7),
respectively. Note that the case with $d=1$ is not considered because $v_{k}$
here exploited to synthesize a view (i.e., at least view $v_{k-1}$).

Specifically, for view $v_{k-d}$, denote %$I_{d}=\set{v_{\min\{m,k-D\}+1},%
%\ldots,v_{\min\{m,k-d\}-1}}$ and
$E_{d}=\{v_{\min\{m,k-d\}+1},\ldots,v_{k-1}\}$ %\emph{
%internal-exploration view set} and
, %respectively,
where all views in $E_{d}$ are forced to select $(v_{k-d}, v_{k})$ for synthesis with DIBR.
This is because when $k-d$ is the maximum index (other than $k$) such that
$v_{k-d}$ is transmitted directly in $\theta^*_{m,k}$,
no views between $v_{k-d}$ and $v_k$ can transmitted directly
and thus must select $(v_{k-d}, v_{k})$ for synthesis with DIBR,
for otherwise it will create crossing view selections, which
is forbidden in the definition of the MMDS problem.
Fig.~\ref{fig:exploreration-stage} presents an illustrative example.
Therefore, it is necessary to multicast view $v_{k-d}$ to not only the
clients subscribing view $v_{k-d}$ but also all the other clients
subscribing the views in $E_{d}$.
%Most importantly, $v_{k-d}$ may also be
%exploited to synthesize any other view in $I_{d}$. In other words, the
%bandwidth cost of the multicast tree to deliver $v_{k-d}$ varies by the
%clients in $I_{d}$ that exploit $v_{k-d}$ for synthesis, since the clients
%subscribing $v_{k-d}$ and $E_{d}$ all need to receive $v_{k-d}$. Therefore,
%it is necessary to carefully examine the view selection for $I_{d}$. \textbf{%
%(Please read the above sentences carefully.)}

%Let $\Theta (v)$ denote the \emph{view-selection set} of view $v$, which
%represents the set of all possible selections of view $v$ (i.e., directly
%transmitting $v$ or synthesizing $v$ with possible left and right views).
%Let the set of all possible view-selection combinations of the views in $%
%I_{d}$ when $v_{k-d}$ is explored is denoted by $\Gamma _{d}$. In other
%words, $\Gamma _{d}$ is the collection of all possible selection
%combinations for the views in $I_{d}$. Each $\Gamma $ in $\Gamma _{d}$ is
%called a \emph{view combination}. Some views in $I_{d}$ are transmitted
%directly, and the others are synthesized with DIBR. Therefore, a view is
%called a \emph{fixed view} if it is transmitted directly in a
%view-combination $\Gamma \in \Gamma _{d}$. For each view combination $\Gamma
%\in \Gamma _{d}$, the set of fixed views is denoted by $F(d,\Gamma )$. In
%other words, $F(d,\Gamma )$ includes the multicasted views from $v_{k-D}$ to
%$v_{k-d}$.

For $d\in\{1,2,\ldots,\min\{m,k-D\}\}$,
let $c_{m,k}^{d}$ denote the bandwidth consumption to serve the clients
with the desired views from $v_{m}$ to $v_{k}$, where $v_{k}$ is employed to
synthesize for all the views from $v_{k-1}$ to $v_{k-d}$.
MMDEA computes and store $c_{m,k}^d$ sequentially for $d=0,1,2,\ldots,\min\{m,k-D\}-1$
according to %the following two cases.
%1)~If $F(d,\Gamma )\cap \{v_{k-D},\ldots ,v_{k-d}\}=v_{k-d}$, i.e., $\Gamma $
%contains only $v_{k-d}$ as the fixed views, the value of $c_{m,k}^{\Gamma }$
%can be derived according to
$c_{m,k-d}^{j}$, $c_{k}$ and $\sum_{v\in E_{d}}\Phi _{\{v_{k-d},v_k\}}^{\{v\}}$,
where $j\in J=\{0,1,2,\ldots,\min\{m,k-D\}-1\}$.
%$\Gamma ^{\prime }$ in the
%computation of $c_{m,k}$.
In other words, $c_{m,k}^{d}$ is obtained by
looking up the previous derived values $c_{m,k-d}^{j}$ and $%
c_{k}$, together with the expansion cost, where each view $v$ %in $I_{d}$
%selects $\theta (v).\ell $ and $\theta (v).r$ for synthesis with DIBR with $%
%\theta (v)\in \Gamma $, and each view $v$
in $E_{d}$ selects $v_{k-d}$ and $%
v_{k}$ for synthesis. %2)~If $F(d,\Gamma )\cap \{v_{k-D},\ldots
%,v_{k-d}\}\neq v_{k-d}$, i.e., $\Gamma $ contains at least one additional
%fixed view, the value of $c_{m,k}^{\Gamma }$ can be derived according to $%
%c_{m,k-d}^{F^{\prime }}$, $c_{k}$ and $\sum_{v\in I_{d}\cup E_{d},\theta
%(v)\in \Gamma }\Phi _{\{\theta (v).\ell ,\theta (v).r\}}^{\{v\}}$ for all
%possible $F^{\prime }$ in the computation of $c_{m,k}$, \textbf{(I did not
%capture why we need an additional notation }$F^{\prime }$\textbf{, instead
%of directly using }$\Gamma $). where $c_{m,k-d}^{F^{\prime }}$ is defined
%similarly to $c_{m,k}$ with the additional restrictions that the fixed views
%in $F^{\prime }$ must be transmitted directly. In other words, $%
%c_{m,k-d}^{F^{\prime }}$ is the minimum total bandwidth consumption to serve
%all clients subscribing views from $v_{m}$ to $v_{k-d}$ such that two
%boundary views $v_{m}$, $v_{k-d}$ and all views in $F^{\prime }$ must be
%transmitted directly.
The corresponding view-selection function for $c_{m,k}^d$ is denoted by $\theta_{m,k}^d$,
and will be stored in the set $\overline{\theta}_{m,k}$ for further reference.
After finding $c_{m,k}^{d}$ for all $%
k=m,m+1,\ldots ,M$, the minimum cost $c_{m,k}$ is derived by the minimum of $%
c_{m,k}^{0}$ and $c_{m,k}^{d}$ for all possible %$\Gamma \in \Gamma _{d}
%$, where
$d\in\{1,2,\ldots ,\min\{D,k-m\}-1\}$. %Detailed pseudocode is presented
%in \cite{CORR}.

\subsection{Example}

In this section, we demonstrate the computation of the minimum total
bandwidth consumption in Fig.~\ref{fig:MMDS} using MMDEA under $D=4$.
The set of desired views is $\mathcal{V}_\rho=\set{v_m=v_2,v_3,v_4,v_6,v_7,v_8=v_M}$.
Since the gap in $\mathcal{V}_\rho$ is no larger than $D$,
only one segment needs to consider.
Initially, $c_{2,2}=7$.
Afterwards, $c_{2,3}$ must be obtained
by the view-combination that do not involve $v_3$, i.e.,
$c_{2,3}=c_{2,3}^0=c_{2,2}+c_3=14$.
%Also the corresponding view-selection function is
%$\theta_{2,3}: v_2\mapsto(v_2,v_2)$, $v_3\mapsto(v_3,v_3)$,
%and it will be stored in the set $\overline{\theta}_{2,3}$ for further reference.
Now consider $c_{2,4}$. We have $J=\{0,2\}$.
Firstly, $c_{2,4}^0=c_{2,3}+c_4=21$.
In the exploration stage, $d=2$ and we get $E_d=\{v_3\}$.
So we obtain $c_{2,4}^{2}=\min_{j\in J}\{c_{2,2}^j+c_4+\Phi^{\{v_3\}}_{(v_2,v_4)}\}=17$.
Thus $c_{2,4}=\min\{c_{2,4}^0,c_{2,4}^2\}=17$.
The corresponding assignments of $c_{2,4}^0$ and $c_{2,4}^2$
will be stored in the set
$\overline{\theta}_{2,4}=\set{\theta_{2,4}^0,\theta_{2,4}^2}$
for further reference, where
$\theta_{2,4}^0\colon v_2\mapsto(v_2,v_2)$,
$v_3\mapsto(v_3,v_3)$, $v_4\mapsto(v_4,v_4)$, and
$\theta_{2,4}^2\colon v_2\mapsto(v_2,v_2)$, $v_3\mapsto(v_2,v_4)$,
$v_4\mapsto(v_4,v_4)$, respectively.
Next, consider $c_{2,5}$. We have $J=\{0,2,3\}$.
Firstly, $c_{2,5}^0=c_{2,4}+c_5=\infty$
as $v_5\not\in\mathcal{V}_\rho$ and $v_5$ is not generatable
by views from $v_2$ to $v_4$ in $c_{2,5}^0$.
In the exploration stage, $d=2$ and $3$.
For $d=2$, we have $E_d=\{v_4\}$ and
$c_{2,5}^{2}= \min_{j\in J}\{c_{2,3}^{j}+c_5+\Phi^{\{v_4\}}_{(v_3,v_5)}\}=23$.
For $d=3$, we have $E_d=\{v_3,v_4\}$ and $%
c_{2,5}^{3}=\min_{j\in J}\{c_{2,2}^{j}+c_5+\Phi^{\{v_3,v_4\}}_{(v_2,v_5)}\}=19$.
So $c_{2,5}$ is the minimum among $c_{2,5}^0$, $c_{2,5}^{2}$, and $c_{2,5}^{3}$,
which results in $c_{2,5}=19$.
%Similarly, for $c_{2,6}$, $d=2,3$ and $4$.
%Firstly, $c_{2,6}^0 = c_{2,5}+c_6=25$.
%The minimum values of $c_{2,6}^d $ for each $d$ are:
%$c_{2,6}^{\Gamma_1}=c_{2,4}^{\{v_2\}\setminus\{v_3\}}+c_6+\Phi^{v_3}_{(v_2,v_4)}+\Phi^{v_5}_{(v_2,v_6)}=23$;
%$c_{2,6}^{\Gamma_2}=c_{2,4}^{\{v_2\}\setminus\{v_3\}}+c_6+\Phi^{v_3}_{(v_2,v_4)}+\Phi^{v_5}_{(v_4,v_6)}=23$;
%$c_{2,6}^{\Gamma_3}=c_{2,2}^{\{v_2\}}+c_6+\Phi^{\{v_3,v_4,v_5\}}_{(v_2,v_6)}=19$, respectively.
%Thus $c_{2,6}=\min\{c_{2,6}^0,c_{2,6}^{\Gamma_1},c_{2,6}^{\Gamma_2},c_{2,6}^{\Gamma_3}\}=19$.
Similarly, $c_{2,6}=19$.
The value of $c_{2,7}$ can be obtained similarly as
$c_{2,7}=\min_{j\in J}\{c_{2,3}^{j}+c_7+\Phi^{\{v_3,v_4,v_5\}}_{(v_2,v_6)}\}=28$,
where $J=\{0,2,3,4\}$.
%and the corresponding view-selection function $\theta_{2,7}$ is $v_2\mapsto(v_2,v_2)$,
%$v_3\mapsto(v_2,v_6)$, $v_4\mapsto(v_2,v_6)$, $v_6\mapsto(v_6,v_6)$ and
%$v_7\mapsto(v_7,v_7)$.
The value of $c_{2,8}$ can be obtained similarly as
$c_{2,8}=\min_{j\in J}\{c_{2,4}^{j}+c_8+\Phi^{\{v_5,v_6,v_7\}}_{(v_4,v_8)}\}=17+15=32$,
where $J=\{0,2,3,4\}$,
and the corresponding view-selection function $\theta_{2,8}$ is $v_2\mapsto(v_2,v_2)$,
$v_3\mapsto(v_2,v_4)$, $v_4\mapsto(v_4,v_4)$, $v_6\mapsto(v_4,v_8)$,
$v_7\mapsto(v_4,v_8)$ and $v_8\mapsto(v_8,v_8)$.
Consequently, the minimum total bandwidth consumption
with respect to $D=4$ in this example is $c_{m,M}=c_{2,8}=32$.

\subsection{Optimality}

The solution optimality of MMDEA relies on the correctness of $c_{m,k}$ for
all $k\in \set{m,m+1,\ldots,M}$, which can be proved similarly as in Lemma~%
\ref{lem:D2} by induction on $k$. If $v_k$ is not
exploited to synthesize any other view in
$\theta^*_{m,k}$, clearly $c_{m,k}$ is $c_{m,k}^0$; otherwise, the
value $c_{m,k}$ must be obtained by examining all subproblems that must
exploit $v_k$ for synthesis with DIBR. The algorithm checks all possible view
selection combinations for the views from $v_{k-D+1}$ to $v_{k-1}$ as only
these views have the abilities to exploit $v_k$ for synthesis with DIBR.
%%The view selection combinations of the views from $v_{k-D+1}$ to $v_{k-1}$
%%can be achieved by the exploration of $d$ from $2$ to $\min\{D,k-m\}$ systematically,
%%which corresponds to $v_{k-1},v_{k-2},\ldots,v_{\min\{D,k-m\}}$, respectively.
Thus the optimization problem for $v_m,\ldots,v_k$ (i.e., $c_{m,k}$) can be
obtained by looking up the subproblem $v_m,\ldots,v_{k-d}$ (i.e., $c_{m,k-d}$).
%with the restrictions in the view-combination $\Gamma\in\Gamma_d$.
Since $c_{m,k}$ is a minimization, by comparing the optimal solution among $%
c_{m,k}^0$ and $c_{m,k}^{d}$ for all possible $d$% and all possible
%selection combinations $\Gamma\in\Gamma_d$,
the optimal solution $c_{m,k}$ is derived.
%The time complexity of MMDEA is $\mathcal{O}(\cd{V}\cd{\mathcal{V}}D^D)$,
%where $\cd{V}$ is the number of nodes in the network and
%$\cd{\mathcal{V}}$ is the total number of views provided by the server.

\subsection{Time Complexity}
Now we analyze the time complexity of MMDEA. For any $v_k\in\mathcal{V}_\rho$,
the multicast tree for the computation of $c_{k,k}$ can be obtained by
running a tree transversal to identify the edges in $T$ in which the edge
has shortest $s,t$-paths through it for some client user $t\in \Omega_T$
with that $t$ prefers view $v_k$ (i.e., $\rho_T(t)=v_k)$. Similarly, the
multicast tree for the computation of $\Phi_{(v_\ell,v_r)}^{\mathcal{V}%
^\prime}$ for any $\mathcal{V}^\prime \subseteq\mathcal{V}$ can be similarly
determined as $c_{k,k}$. Thus, $c_{k,k}$ and $\Phi_{(v_\ell,v_r)}^{\mathcal{V%
}^\prime}$ can be computed in time $\mathcal{O}(\left|V\right| )$.

The initialization stage and the union stage clearly takes $\mathcal{O}%
(\left|\mathcal{V}\right| )$ time to complete. The time complexity of MMEDA
clearly bound by the time in the exploration stage. For a fixed $d$, there
are up to $D$ %$(D-d+1)^{d-1}$ and $D^{2(D-d-1)}$
possible choices for the views
in %$I_d$ and
$E_d$.%, respectively.
%Thus there are up to $\mathcal{O}%
%\left(\sum_{d=1}^{D-1}(D-d+1)^{d-1}D^{2(D-d-1)}\right)=\mathcal{O}(D^{2D})$
%possible view-combinations to consider for each $k$.
The computed cost $c_{m,k}$ will be stored for further reference.%, and
%thus it take $\mathcal{O}(D^{2D})$ time to find $c_{m,k}^{d}$.
The time complexity of MMDEA is $\mathcal{O}(\cd{V}\cd{\mathcal{V}}D^D)$,
where $\cd{V}$ is the number of nodes in the network and
$\cd{\mathcal{V}}$ is the total number of views provided by the server.%
%Consequently, MMEDA takes time $\mathcal{O}(\left|V\right| \left|\mathcal{V}%
%\right| D^{D}) $.

\section{Extension}

In this section, we consider a generalization
of the MMDS problem which allows crossing-view selections, i.e.,
the views in $\mathcal{V}_\rho$ can select views for synthesis
with DIBR that may create interlacing view selections.
For example, if view $v$ selects $(\theta(v).\ell,\theta(v).r)$
for synthesis with DIBR, for another view $v'$,
$v'$ can select $(\theta(v').\ell,\theta(v').r)$
with $\theta(v).\ell< \theta(v').\ell< \theta(v).r$ or
$\theta(v).\ell< \theta(v').r< \theta(v).r$.
For convenience, we call such extension
the \emph{E-MMDS} problem.
We proposed an algorithm called E-MMDEA to deal with
the E-MMDS problem.

%-----------------------------------------------------------------------------
\DecMargin{1em}
\begin{algorithm}[!t]
\label{algo:MMDEA}
\SetNlSty{}{}{:}
\SetNlSkip{0.05em}
\SetEndCharOfAlgoLine{;}
%\scriptsize
\caption{Multi-view and Multicast Delivery Exploration Algorithm (\textsc{E-MMDEA})}
\KwIn{A rooted tree $T=(V,A,s)$, a universal view set $\mathcal{V}$,
a preferred-view function $\rho_T$, and the DIBR quality constraint $D$.}
\KwOut{The minimum total bandwidth consumption $cost(\theta^*)$ of a view-selection function $\theta^*$
to the E-MMDS problem.}
\KwMD{}
%\scriptsize
\tcp*[h]{Initialization stage}\\
Identify the service range $\mathcal{V}_\rho\gets \mathcal{V}_\rho^1\cup\cdots\cup\mathcal{V}_\rho^n$\;
$cost(\theta^*)\gets 0$\;
\tcp*[h]{Exploration stage}\\
\ForEach{segment $\mathcal{V}_\rho^i \gets \{v_m,\ldots,v_M\}$}{
    \For{$k=m$ \KwTo $M$}{
        $c_{m,k}^0\gets \min\{c_{m,k-1},c_{m,k-2},\ldots,c_{m,k-D}\}+c_k$\;
        %\Indentp{-1.2em}
        $J\gets\{0\}$\;
        \For{$d=2$ \KwTo $\min\{D,k-m\}$}{
            %\Indentp{-1em}
            $I_d\gets \set{v_{\min\{m,k-D\}+1},\ldots,v_{\min\{m,k-d\}-1}}$\;
            $E_d\gets \set{v_{\min\{m,k-d\}+1},\ldots,v_{k-1}}$\;
            \ForEach{$v\in \left(I_d\cup E_d\right)\cap \mathcal{V}_\rho$}{
                %\Indentp{-1.2em}
               % \scalebox{0.7}{
               \tcp*[h]{\scalebox{0.75}{The possible selections for view $v\in I_d$}}\\
                $\Theta(v)\gets \{\theta(v)=(v_\ell,v_r) \mid r-\ell\leq D,$\\
                \hspace{1.2cm} $r\leq k,v_m\leq v_\ell\leq v\leq v_r,v_r\notin E_d\, \}$\;
                %}
            }
            $\Gamma_d\gets$ \begin{minipage}[t]{6cm}{all possible view selection combinations by the views
            in $I_d$ or $E_d$ such that each selection combination satisfies $\mathcal{V}_\rho$ w.r.t. $D$;}\end{minipage}
            $J\gets J\cup \{\Gamma\mid \Gamma\in\Gamma_d\}$\;
%            $c_{m,k}^d \gets \min_{j\in J}\Big\{c^j_{m,k-d} + c_k $\\
%                 %\hspace{1.25cm}$+\sum_{v\in I_d,\theta(v)\in\Gamma}\Phi^{\{v\}}_{\{\theta(v).\ell,\theta(v).r\}}$\\
%                 \hspace{1.25cm}$+\sum_{v\in E_d}\Phi^{\{v\}}_{\{v_{k-d},v_k\}}\Big\}$\;
%            $\theta_{m,k}^d \gets \theta_{m,k-d}^j$\\
%            \hspace{1.2cm}$\cup \set{v\mapsto (v_{k-d},v_k)\mid v\in E_d\cap\mathcal{V}_\rho}$\\
%            \hspace{1.2cm}$\cup \set{v_k\mapsto (v_k,v_k)\mid v_k\in \mathcal{V}_\rho}$\;
%
%        }
%        $\overline{\theta}_{m,k}\gets \bigcup_{d=0}^{\min\{D,m-k\}-1} \theta_{m,k}^d$\;
        }
        \ForEach{$d$  {\bf and} $\Gamma\in \Gamma_d, $}{
            \If{$F(d,\Gamma)\cap \{v_{k-D},\ldots,v_{k-d}\}=v_{k-d}$}{
                $c_{m,k}^\Gamma \gets \min_{j\in J}\Big\{c^j_{m,k-d} + c_k $\\
                 \hspace{1.25cm}$+\sum_{v\in I_d,\theta(v)\in\Gamma}\Phi^{\{v\}}_{\{\theta(v).\ell,\theta(v).r\}}$\\
                 \hspace{1.25cm}$+\sum_{v\in E_d}\Phi^{\{v\}}_{\{v_{k-d},v_k\}}\Big\}$\;
            }
            \Else{
                $c_{m,k}^\Gamma \gets \min_{F\in \mathcal{F}}\Big\{c^F_{m,k-d} + c_k $\\
                 \hspace{1.25cm}$+\sum_{v\in I_d,\theta(v)\in\Gamma}\Phi^{\{v\}}_{\{\theta(v).\ell,\theta(v).r\}}$\\
                 \hspace{1.25cm}$+\sum_{v\in E_d}\Phi^{\{v\}}_{\{v_{k-d},v_k\}}\Big\}$\;
            }
        }
        $\theta_{m,k}^0\gets \theta_{m,k-1}\oplus \set{v_k\mapsto (v_k,v_k)\mid \hbox{if }v_k\in\mathcal{V}_\rho}$\;
        $\overline{\theta}_{m,k}\gets \overline{\theta}_{m,k} \bigcup \theta_{m,k}^0$\;

        %$\theta_{m,k}\gets \arg\min c_{m,k}$\;
        $\theta_{m,k}^*\gets \theta_{m,k}^d$ with $c_{m,k}^d=c_{m,k}$\;
        $\overline{\theta}_{m,k}\gets \bigcup_d \overline{\theta}_{m,k}^d$\;
%        $c_{m,k}\gets \min\Big\{c_{m,k}^0,c_{m,k}^{d}\mid $\\
%        \hspace{1.2cm} $d\in\set{1,2,\ldots,\min\{D,k-m\}-1}\Big\}$\;
    }
    %$\theta_i^*\gets \theta_{m,M}$\;
    $cost(\theta^*)\gets cost(\theta^*)+c_{m,M}$\;
}
\Return $cost(\theta^*)$\;
\end{algorithm}

The notion of exploration stage goes as follows. If $v_{k}$ is not exploited
to synthesize any other view, clearly $c_{m,k}=c_{m,k}^{0}=\min
\{c_{m,k-1},c_{m,k-2},\ldots ,c_{m,k-D}\}+c_{k}$, such as Eq. (5) for $D=3$.
Otherwise, it is necessary to examine different view selection combinations
that exploit $v_{k}$ for synthesis with DIBR. To find $c_{m,k}$ in this
case, MMDEA sequentially examines the case that a view $v_{k-d}$, is
transmitted, where $D\geq d\geq 2$. In addition, every other view between $%
v_{k-d}$ and $v_{k}$ is synthesized from the two views accordingly. For
example, when $D=3$, $v_{k-2}$ and $v_{k-3}$ are examined sequentially and
assumed to be transmitted, as explained in Eq. (6) and Eq. (7),
respectively. Note that the case with $d=1$ is not considered because $v_{k}$
here exploited to synthesize a view (i.e., at least view $v_{k-1}$).

Specifically, for view $v_{k-d}$, denote $I_{d}=\set{v_{\min\{m,k-D\}+1},%
\ldots,v_{\min\{m,k-d\}-1}}$ and
$E_{d}=\{v_{\min\{m,k-d\}+1},\ldots,v_{k-1}\}$ %\emph{
%internal-exploration view set} and
, %respectively,
where all views in $E_{d}$ are forced to be synthesized with DIBR.
%This is because when $k-d$ is the maximum index (other than $k$) such that
%$v_{k-d}$ is transmitted directly in $\theta^*_{m,k}$,
%no views between $v_{k-d}$ and $v_k$ can transmitted directly
%and thus must select $(v_{k-d}, v_{k})$ for synthesis with DIBR,
%for otherwise it will create crossing view selections, which
%is forbidden in the definition of the MMDS problem.
%Fig.~\ref{fig:exploreration-stage} presents an illustrative example.
Therefore, it is necessary to multicast view $v_{k-d}$ to not only the
clients subscribing view $v_{k-d}$ but also all the other clients
subscribing the views in $E_{d}$.
Most importantly, $v_{k-d}$ may also be
exploited to synthesize any other view in $I_{d}$. In other words, the
bandwidth cost of the multicast tree to deliver $v_{k-d}$ varies by the
clients in $I_{d}$ that exploit $v_{k-d}$ for synthesis, since the clients
subscribing $v_{k-d}$ and $E_{d}$ all need to receive $v_{k-d}$. Therefore,
it is necessary to carefully examine the view selection for $I_{d}$.

Let $\Theta (v)$ denote the \emph{view-selection set} of view $v$, which
represents the set of all possible selections of view $v$ (i.e., directly
transmitting $v$ or synthesizing $v$ with possible left and right views).
Let the set of all possible view-selection combinations of the views in $%
I_{d}$ when $v_{k-d}$ is explored is denoted by $\Gamma _{d}$. In other
words, $\Gamma _{d}$ is the collection of all possible selection
combinations for the views in $I_{d}$. Each $\Gamma $ in $\Gamma _{d}$ is
called a \emph{view combination}. Some views in $I_{d}$ are transmitted
directly, and the others are synthesized with DIBR. Therefore, a view is
called a \emph{fixed view} if it is transmitted directly in a
view-combination $\Gamma \in \Gamma _{d}$. For each view combination $\Gamma
\in \Gamma _{d}$, the set of fixed views is denoted by $F(d,\Gamma )$. In
other words, $F(d,\Gamma )$ includes the multicasted views from $v_{k-D}$ to
$v_{k-d}$.

For $d\in\{1,2,\ldots,\min\{m,k-D\}\}$,
let $c_{m,k}^{d}$ denote the bandwidth consumption to serve the clients
with the desired views from $v_{m}$ to $v_{k}$, where $v_{k}$ is employed to
synthesize for all the views from $v_{k-1}$ to $v_{k-d}$.
MMDEA computes and store $c_{m,k}^d$ sequentially for $d=0,1,2,\ldots,\min\{m,k-D\}-1$
according to the following two cases.
1)~If $F(d,\Gamma )\cap \{v_{k-D},\ldots ,v_{k-d}\}=v_{k-d}$, i.e., $\Gamma $
contains only $v_{k-d}$ as the fixed views, the value of $c_{m,k}^{\Gamma }$
can be derived according to
$c_{m,k-d}^{j}$, $c_{k}$ and $\sum_{v\in E_{d}}\Phi _{\{v_{k-d},v_k\}}^{\{v\}}$,
where $j\in J=\{0,1,2,\ldots,\min\{m,k-D\}-1\}$.
$\Gamma ^{\prime }$ in the
computation of $c_{m,k}$.
In other words, $c_{m,k}^{d}$ is obtained by
looking up the previous derived values $c_{m,k-d}^{j}$ and $%
c_{k}$, together with the expansion cost, where each view $v$ %in $I_{d}$
selects $\theta (v).\ell $ and $\theta (v).r$ for synthesis with DIBR with $%
\theta (v)\in \Gamma $, and each view $v$
in $E_{d}$ selects $v_{k-d}$ and $%
v_{k}$ for synthesis. 2)~If $F(d,\Gamma )\cap \{v_{k-D},\ldots
,v_{k-d}\}\neq v_{k-d}$, i.e., $\Gamma $ contains at least one additional
fixed view, the value of $c_{m,k}^{\Gamma }$ can be derived according to $%
c_{m,k-d}^{F^{\prime }}$, $c_{k}$ and $\sum_{v\in I_{d}\cup E_{d},\theta
(v)\in \Gamma }\Phi _{\{\theta (v).\ell ,\theta (v).r\}}^{\{v\}}$ for all
possible $F^{\prime }$ in the computation of $c_{m,k}$, $F^{\prime }$,
where $c_{m,k-d}^{F^{\prime }}$ is defined
similarly to $c_{m,k}$ with the additional restrictions that the fixed views
in $F^{\prime }$ must be transmitted directly. In other words, $%
c_{m,k-d}^{F^{\prime }}$ is the minimum total bandwidth consumption to serve
all clients subscribing views from $v_{m}$ to $v_{k-d}$ such that two
boundary views $v_{m}$, $v_{k-d}$ and all views in $F^{\prime }$ must be
transmitted directly.
The corresponding view-selection function for $c_{m,k}^d$ is denoted by $\theta_{m,k}^d$,
and will be stored in the set $\overline{\theta}_{m,k}$ for further reference.
After finding $c_{m,k}^{d}$ for all $%
k=m,m+1,\ldots ,M$, the minimum cost $c_{m,k}$ is derived by the minimum of $%
c_{m,k}^{0}$ and $c_{m,k}^{d}$ for all possible $\Gamma \in \Gamma _{d}
$, where
$d\in\{1,2,\ldots ,\min\{D,k-m\}-1\}$. %Detailed pseudocode is presented
%in \cite{CORR}.

\section{Heuristic Algorithm Design}

\subsection{Design of H-MMDEA}

Even though MMDEA is able to optimally select optimal views and deliver
optimal multi-view videos over IP networks, the algorithm results in a high
computational cost for the network with large $D$. To address the issue, we
propose a heuristic algorithm called H-MMDEA to acquire the solution in a
linear time. Recall that the complexity of MMDEA comes from two parts.
First, it examines a great number of view transmissions for each service
range. Second, MMDEA is required to determine the view transmission among
the stored possible view transmissions as performing each exploration. To
reduce the complexity, we design H-MMDEA to improve multicast delivery by
iteratively examining alternative transmissions, instead of examining large
number of possible view transmissions for the optimal solution.

H-MMDEA includes three steps: 1) Desired View Setting, 2) Alternative View
Examination, and 3) Multicast Delivery Adjustment. In the fi{}rst step, the
multi-view video server delivers the views directly based on the desired
views clients request. In the second step, the routers in the network
examine alternative view transmission for desired views. In the third step,
the server selects the most efficient alternative view transmission and
adjusts the multicast delivery. H-MMDEA iteratively processes steps 2 and 3
if alternative view transmissions have a better performance. Algorithm 3 details H-MMDEA.

\begin{algorithm}[!t]
\label{alg:DH-MMDEA}
\SetNlSty{}{}{:}
\SetNlSkip{0.05em}
\SetEndCharOfAlgoLine{;}
\caption{\label{alg:DH-MMDEA} Heuristic View and Multicast Delivery Exploration Algorithm (\textsc{H-MMDEA})}
\KwIn{The multicast SPT routing $\mathcal{G}=(\mathcal{V},\mathcal{E})$; request
view $y_{tk}$ for each $t\in\mathcal{T}\textrm{ and }k\mathcal{\in\mathcal{K}}$}
\KwOut{The set of selected
views $\mathcal{M}_{s}$ and multiview video multicast delivery $x_{ijk}$
for each $(i,\: j)\mathcal{\in E}$ and $k\in\mathcal{K}$}
\KwMD{}
1: Obtain a postorder set $\mathcal{\tilde{V}}$ which orders nodes in $\mathcal{G}$.\\
2: Initial setting:
$\mathcal{M}_{i}=\{\}$,$\mathcal{M}_{i}^{v_{k}^{best}}=\{\}$,$u(\mathcal{M}_{i}^{v_{k}^{best}})=\infty$,
$\forall i\in\mathcal{\mathcal{\tilde{V}}}$, $k^{'}=0$, $\mathcal{C}=\{\}$,
$\mathcal{M}_{s}$ and $u(\mathcal{M}_{s})$ can be obtained by directly delivering desired views.\\
\While{$u(\mathcal{M}_{s})<u(\mathcal{M}_{s}^{t*})$}{
\ForEach{$v_{k}\in\mathcal{M}_{s}\setminus\left\{ v_{1},v_{K}\right\}$}{$\mathcal{S}(v_{k})\leftarrow\left\{ \left\{ l,r\right\} \mid r-l\leq D,l\leq k\leq r,l,r\in\mathcal{M}_{s}\right\}$;\\
    \ForEach{$\forall\left\{ l,r\right\} \in\mathcal{S}(k)$}{
        \ForEach{$i\in\mathcal{T}$}{
            \If{$\mathcal{M}_{i}=\{v_{k}\}$}{
                $\mathcal{M}_{i}^{k}=\left\{ l,r\right\}$;}}
        \ForEach{$i\in\mathcal{\tilde{V}\setminus\mathcal{T}}$}{
             $\mathcal{M}_{i}^{k}=\bigcup_{j\mathcal{\in\delta}^{+}(i)}\mathcal{M}_{j}^{k}$;
            \\$u(\mathcal{M}_{i}^{k})=\sum_{\forall j\in\delta^{+}(i)}\left(u(\mathcal{M}_{j}^{k})+|\mathcal{M}_{j}^{k}|\right)$;}
        \If{$u(\mathcal{M}_{s}^{k*})>u(\mathcal{M}_{s}^{k})$}{$\mathcal{M}_{i}^{k*}=\mathcal{M}_{i}^{k},\forall i\in\mathcal{\tilde{V}}$;}
                }}
$t=\arg\min_{k\in\mathcal{M}_{s}\setminus\left\{ 1,K\right\} }\left\{ u\left(\mathcal{M}_{s}^{k*}\right)\right\}$;\\
\If{$u(\mathcal{M}_{s})>u\left(\mathcal{M}_{s}^{t*}\right)$}{$\mathcal{M}_{i}=\mathcal{M}_{s}^{t*},\forall i\in\mathcal{\tilde{V}}$;}
}
\Return
\[
x_{ijk}=\begin{cases}
1 & ,\textrm{ if }v_{k}\in\mathcal{M}_{j}\\
0 & ,\textrm{ otherwise }
\end{cases},\forall(i,\: j)\mathcal{\in E},\forall k\in\mathcal{K};.
\]

\end{algorithm}

\section{Simulations}\label{sec:simulation}

\begin{figure*}[!tp]
\begin{minipage}[t]{1.75 in}
\includegraphics[width=1.75 in,angle=270]{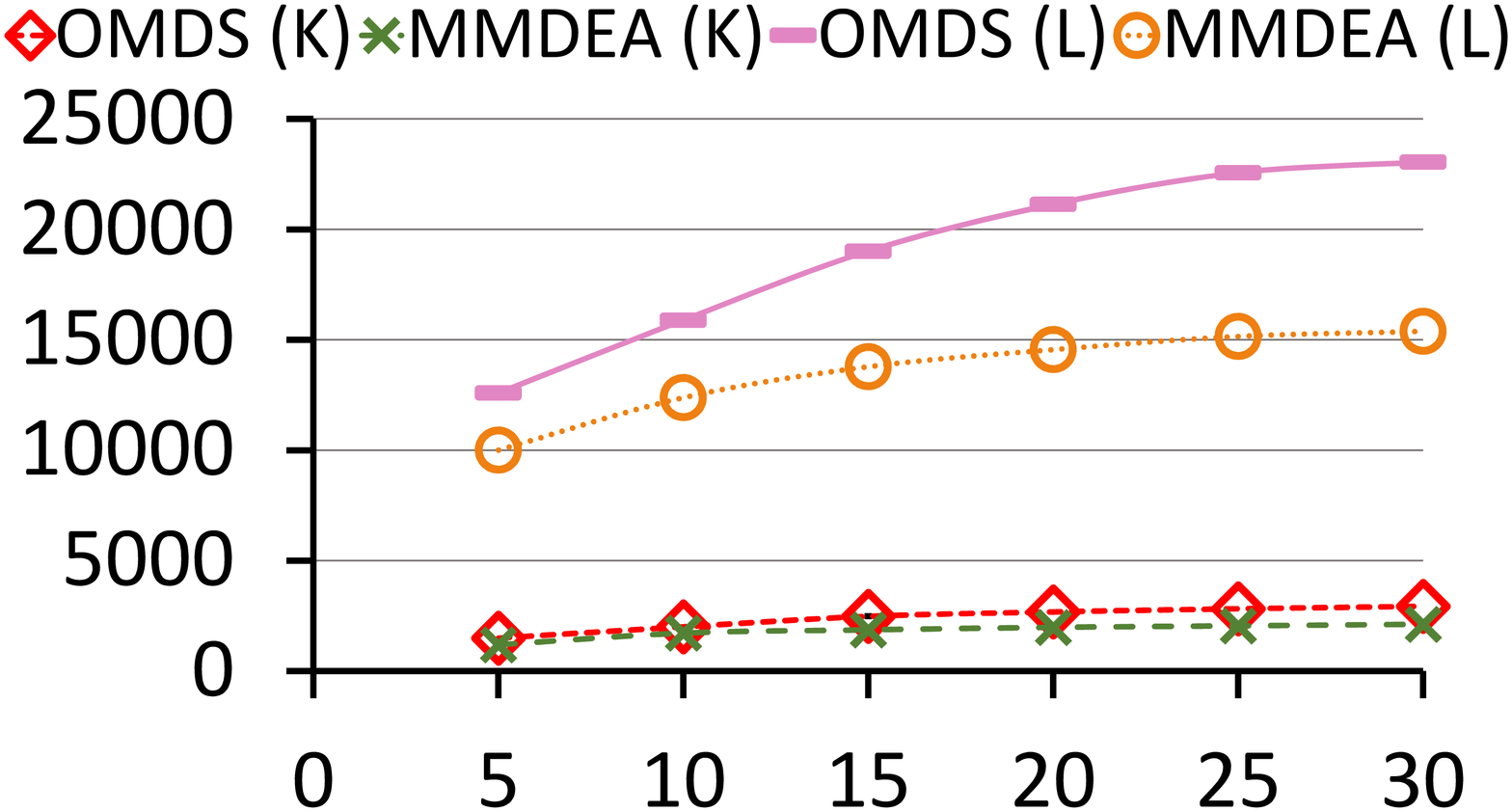}
\vspace{-1.5cm}
\captionsetup{width=1.6in}
\caption{Scenario 1 and 3 (x: number of views, y: total bandwidth consumption)}\label{1}
\end{minipage}
\begin{minipage}[t]{1.75 in}
\includegraphics[width=1.75 in,angle=270]{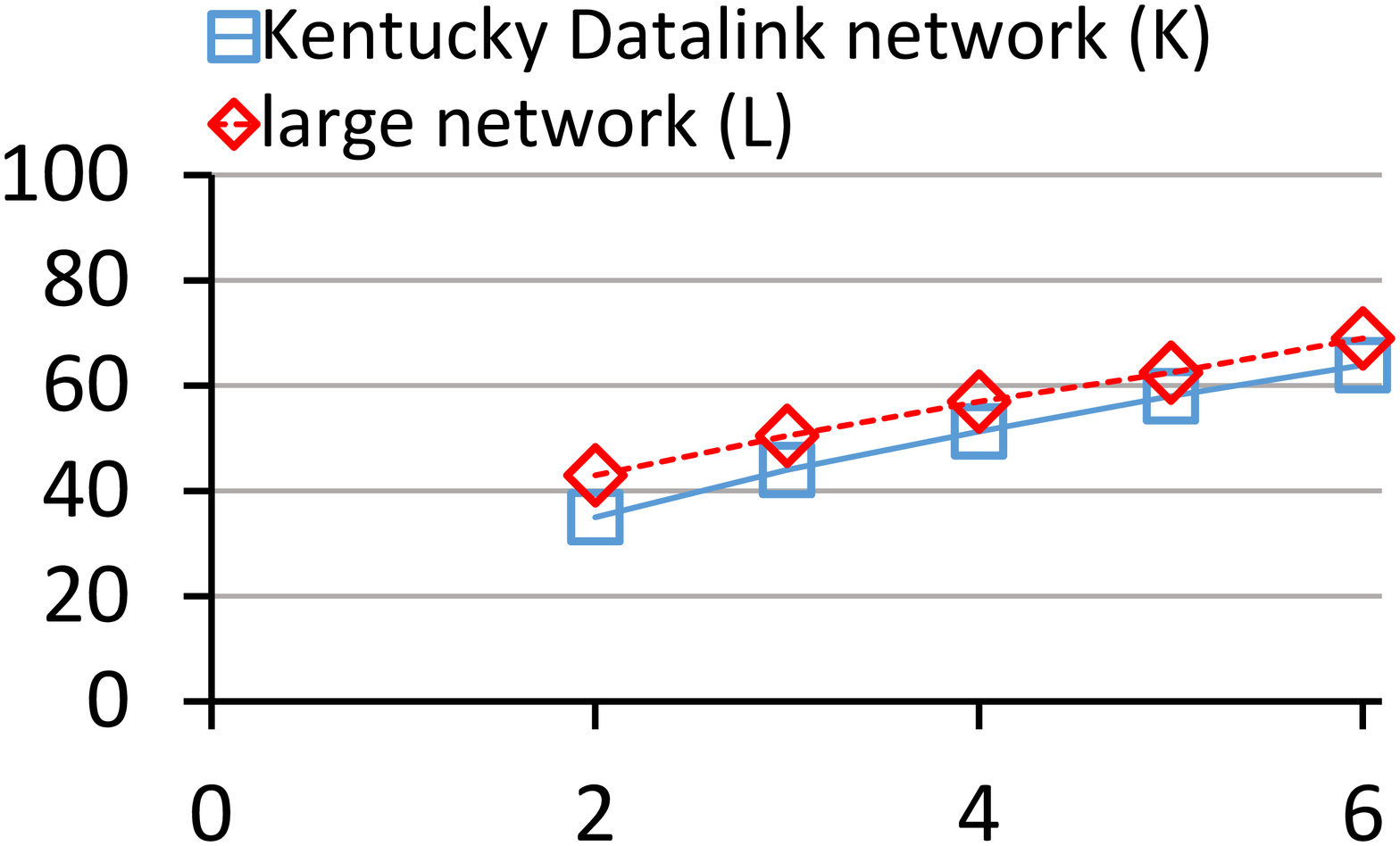}
\vspace{-1.5cm}
\captionsetup{width=1.6in}
\caption{Scenario 1 (x: $D$, y: Percentage of receiving two views.)}\label{1-2}
\end{minipage}
\begin{minipage}[t]{1.75 in}
\includegraphics[width=1.75 in,angle=270]{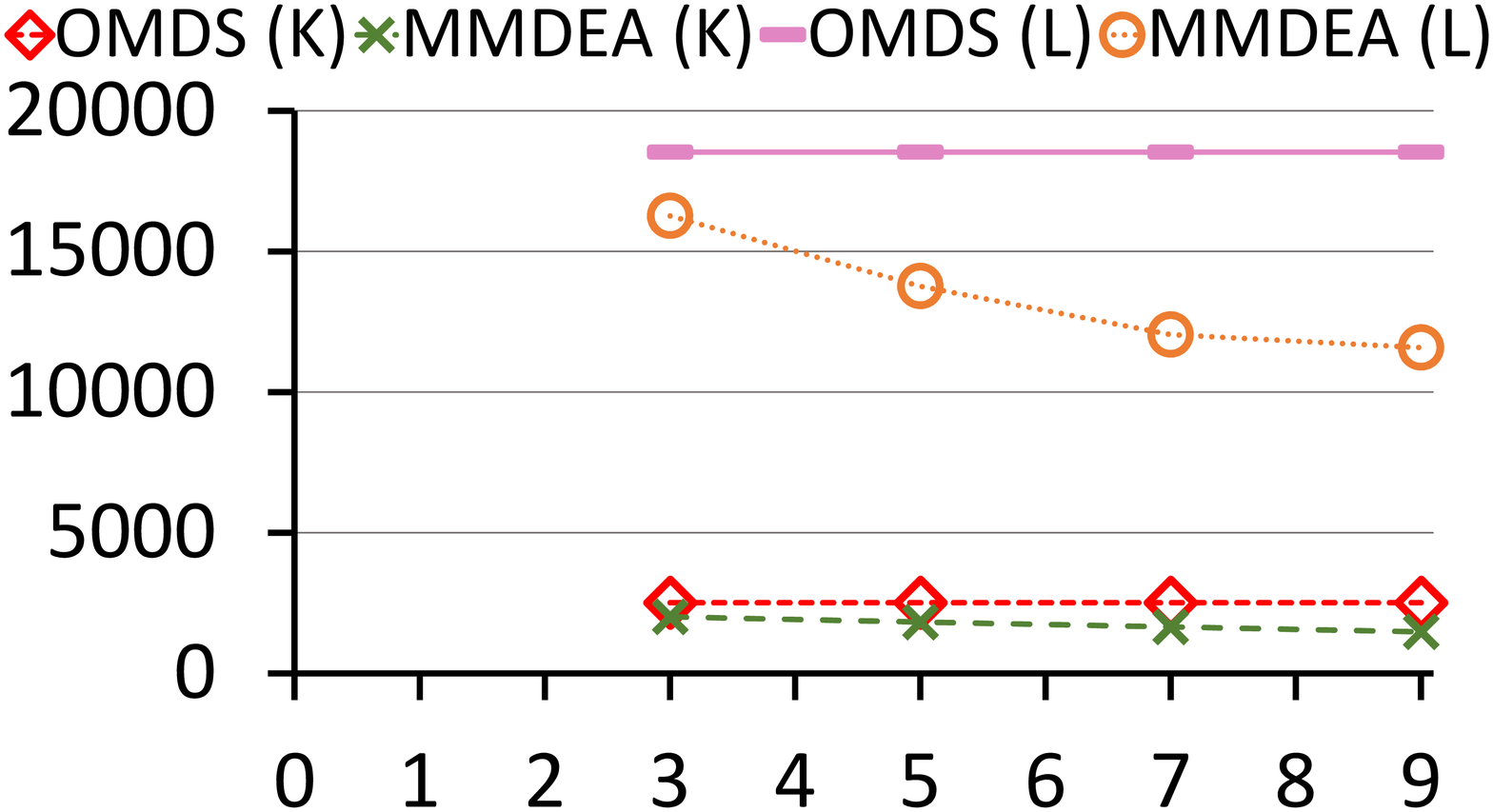}
\vspace{-1.5cm}
\captionsetup{width=1.6in}
\caption{Scenario 2 (x: $D$, y: total bandwidth consumption)}\label{1-1}
\end{minipage}
\begin{minipage}[t]{1.75 in}
\includegraphics[width=1.75 in,angle=270]{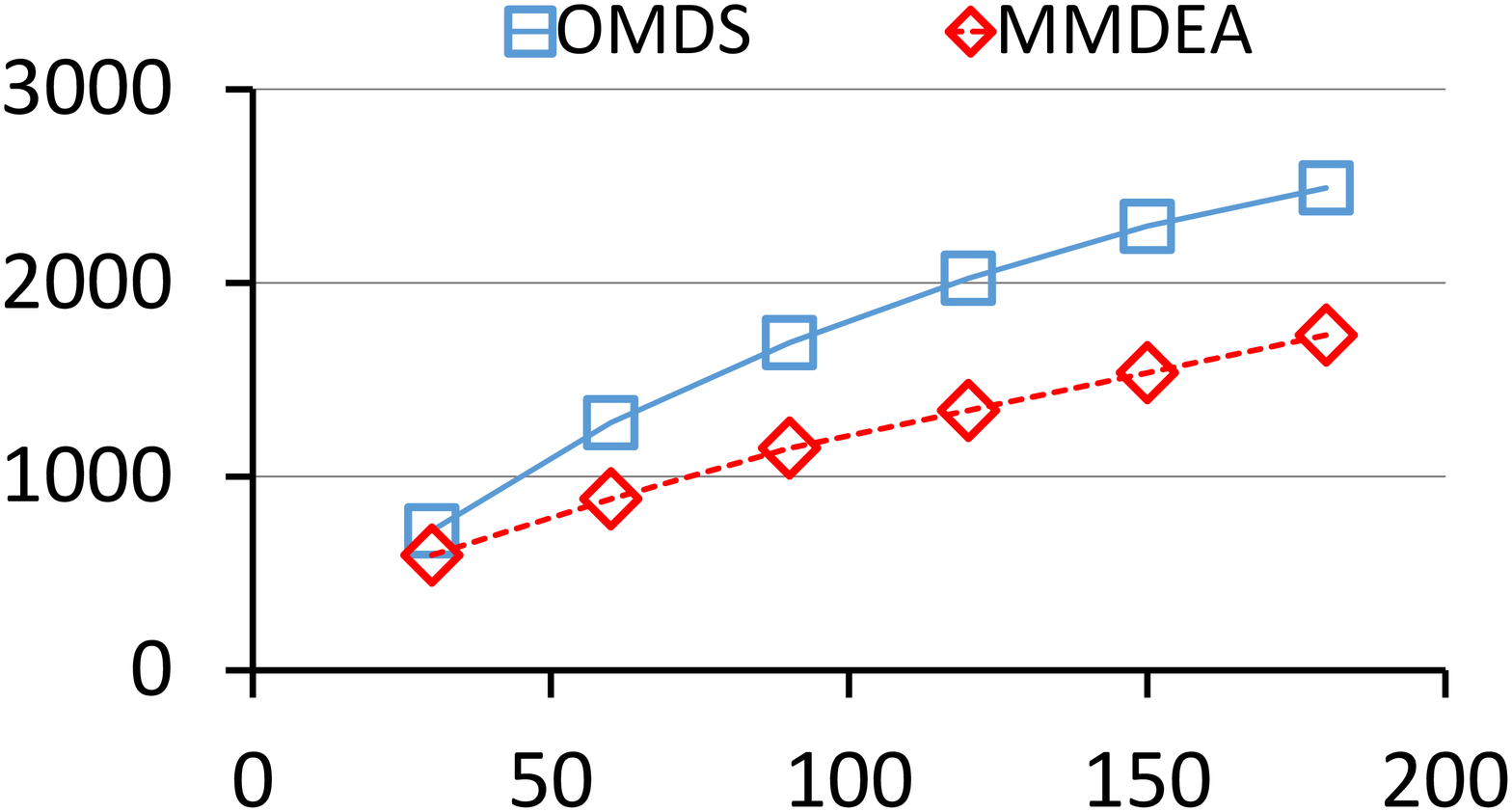}
\vspace{-1.5cm}
\captionsetup{width=1.6in}
\caption{Scenario 4 (x: number of clients, y: total bandwidth consumption)}\label{2-1}
\end{minipage}
\begin{minipage}[t]{1.75 in}
\includegraphics[width=1.75 in,angle=270]{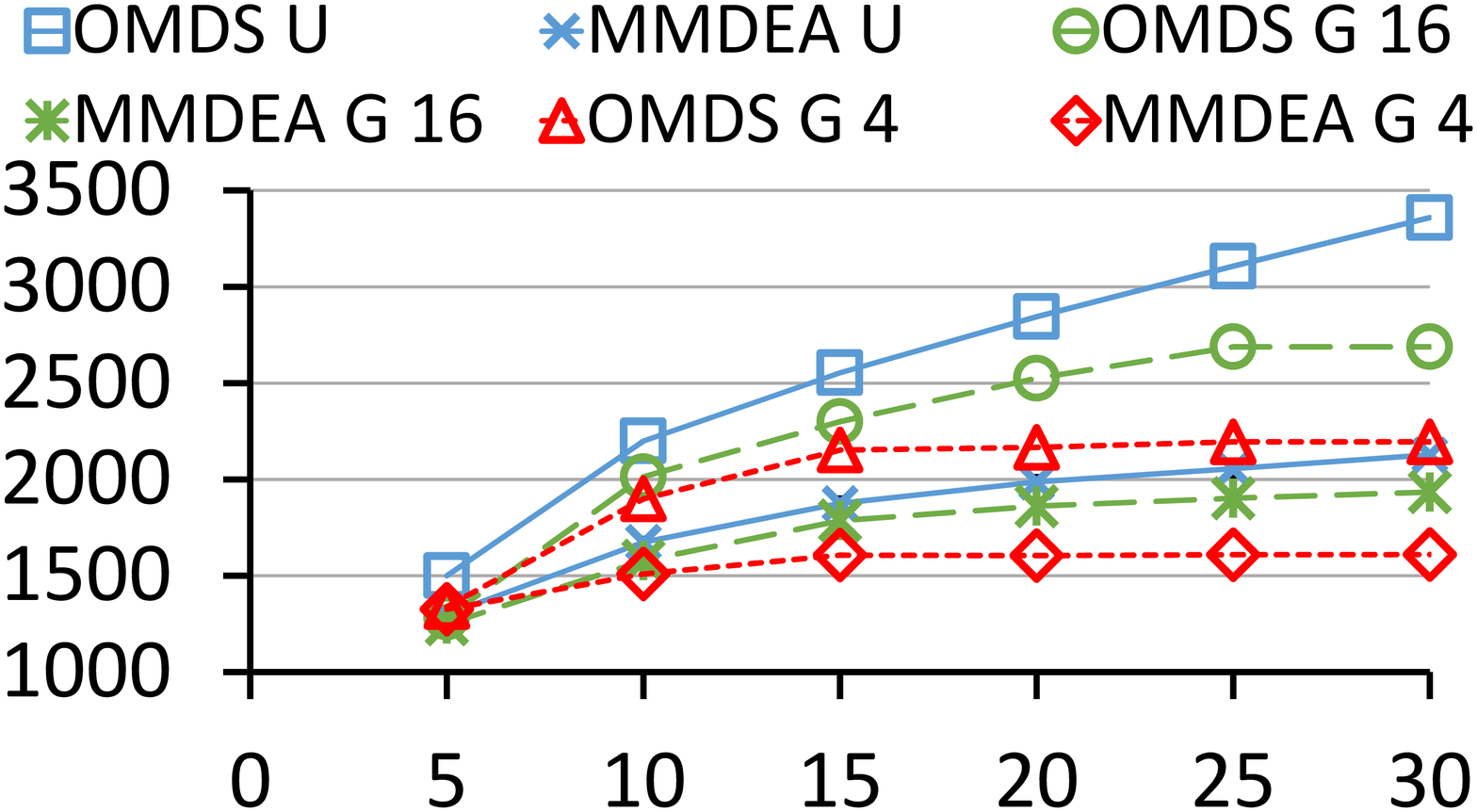}
\vspace{-1.5cm}
\captionsetup{width=1.6in}
\caption{Scenario 5 (x: number of views, y: total bandwidth consumption)}\label{3-1}
\end{minipage}
\begin{minipage}[t]{1.75 in}
\includegraphics[width=1.75 in,angle=270]{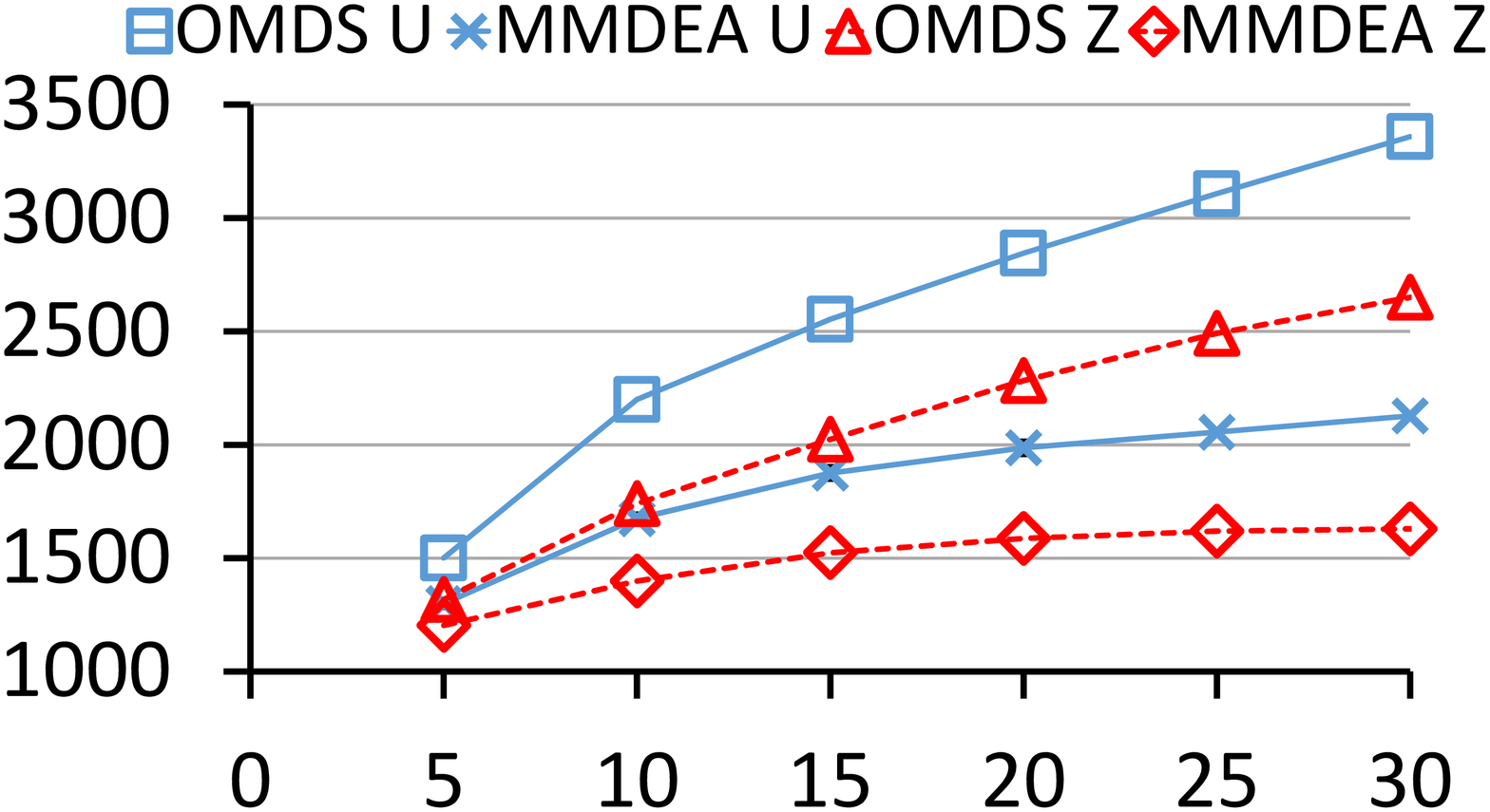}
\vspace{-1.5cm}
\captionsetup{width=1.6in}
\caption{Scenario 5 (x: number of views, y: total bandwidth consumption)}\label{3-2}
\end{minipage}
\begin{minipage}[t]{1.75 in}
\includegraphics[width=1.75 in,angle=270]{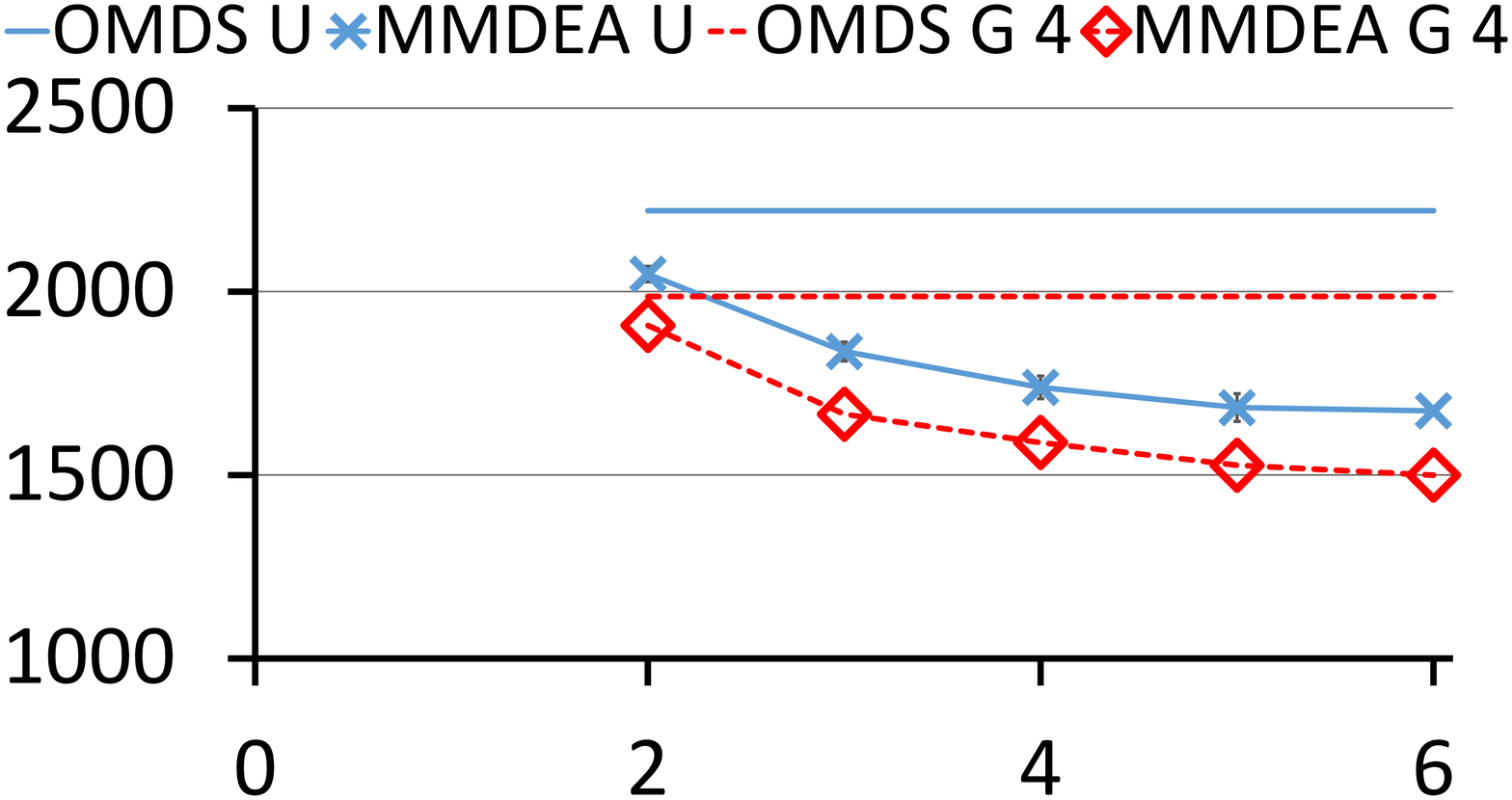}
\vspace{-1.5cm}
\captionsetup{width=1.6in}
\caption{Scenario 5 (x: $D$, y: total bandwidth consumption)}\vspace{12pt}\label{3-3}
\end{minipage}
\begin{minipage}[t]{1.75 in}
\includegraphics[width=1.75 in,angle=270]{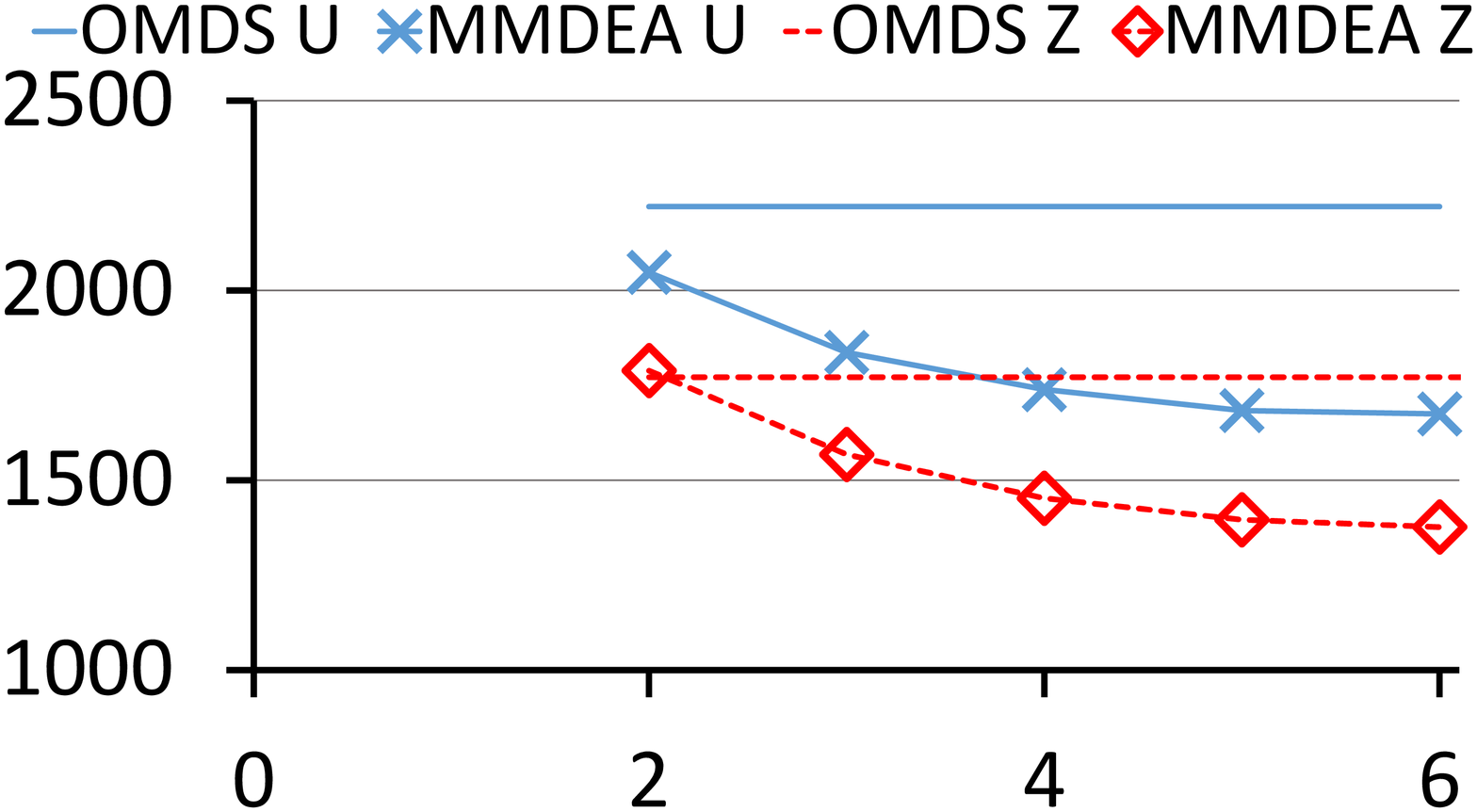}
\vspace{-1.5cm}
\captionsetup{width=1.6in}
\caption{Scenario 5 (x: $D$, y: total bandwidth consumption)}\label{3-4}
\end{minipage}
\end{figure*}

In this section, we compare MMDEA with the existing multicast scheme in a real network \cite{3D-Knight2011} and in the networks generated by Inet \cite{3D-Inet}.

We first conduct the simulation in a small real network called the Kentucky Datalink Network (K) with 754 nodes and 895 links, and a large network (L) with 10000 nodes and 20576 links. We compare MMDEA with the original multicast delivery scheme (OMDS), in which all desired views are multicast separately to the clients without exploiting DIBR. We change the number of views, quality constraint $D$, and the size of networks, i.e, number of clients in the simulation. The performance metrics include the total bandwidth consumption in the network and the
percentage of clients exploiting DIBR to synthesize the desired views. All
algorithms are implemented in an IBM server with four Intel Xeon E7-4820 2.0
GHz CPUs and 48 GB RAM. Each simulation result is averaged over 100 samples.

\subsection{Scenario 1: Size of Networks}

Fig. \ref{1} compares MMDEA with OMDS under the Kentucky Datalink Network (K) and the large network (L) with different numbers of views, where $D$ is $5$. The total bandwidth consumption increases in both schemes with the number of views. Nevertheless, bandwidth consumption for MMDEA is about $35\%$ lower thanks to the efficient aggregation of views with DIBR. More importantly, the improvement becomes more significant when clients are provided with an expanded selection of view. In MMDEA, not all desired views need to be transmitted. As the number of views exceeds $20$, the total bandwidth consumption saturates in both schemes. For OMDS, almost all views are transmitted, while any nearby two transmitted views in MMDEA can be separated with at most $D-1$ views.

Fig. \ref{1-2} shows the percentage of clients receiving two views in the Kentucky Datalink Network (K) and large network (L). The number of views $|\mathcal{V}|$ is set to $12$. When $D$ increases, the percentages of clients synthesizing the desired view in the two networks also grows, which implies that it is not necessary to directly transmit the desired views to all clients since many clients can synthesise their desired views from views subscribed by other clients, thus effectively reducing total bandwidth consumption.

\subsection{Scenario 2: Synthesized range}

Fig. \ref{1-1} evaluates MMDEA with different value of $D$ for the Kentucky Datalink Network (K) and the large network (L) with the number of views set at $12$. The total bandwidth consumption is efficiently reduced as $D$ increases, indicating that it is unnecessary to set a large $D$  because marginal improvement becomes small as $D$ increases, thus indicating that  a  small $D$ (i.e., limited quality degradation) is sufficient to effectively reduce bandwidth consumption in the networks.
\subsection{Scenario 3: Number of views}

Fig. \ref{1} shows the impact of DIBR on different numbers
of views in a video. The bandwidth consumption in both schemes  increase as the video contains more views. The reason is that more views need to be transmitted since desired view of each client follows the uniform distribution. Nevertheless, the result manifests that MMDEA consistently outperforms the OMDS for varied numbers of views.

\subsection{Scenario 4: Number of clients}

Fig. \ref{2-1} shows that the total bandwidth consumption increases  in  both  schemes  with  more  clients. Performance is evaluated under the Kentucky Datalink Network. $|\mathcal{V}|$ and $D$ are respectively set to $12$ and $5$.
Nevertheless, MMDEA achieves an improvement of about $50\%$ thanks to the efficient aggregation of views with DIBR. More
importantly, it is worth noting that the improvement becomes more significant with
more clients in the network because it is easier to find a nearby client that subscribes to a close left view or right view, thus increasing the chance to leverage DIBR.

\subsection{Scenario 5: Distribution of client preferences}

Figs. \ref{3-1} and \ref{3-2} examine the impact of the  distributions of the preferred views. Performance is evaluated using the Kentucky Datalink Network, and the desired views follow the Uniform distribution (U), Gaussian distribution (G) and Zipf distribution (Z) in this scenario. The Zipf distribution is written as $f(l;s;|\mathcal{V}|)=(1/l^2)/\sum_{n=1}^|\mathcal{V}|(1/n^s)$, where $l$ is the preference rank of a view, $s$ is the value of the exponent characterizing the  distribution, and $|\mathcal{V}|$ is the number of views. We set $s=2$ and $|\mathcal{V}|=12$ in the Zipf distribution, which means that clients prefer subscribing only a few important views. In the Gaussian distribution, the smaller variance represents that the desired views of clients are more concentrated. The mean is set at $0.5|\mathcal{V}|$, and the variance is set at $4$ and $16$ in this paper. The result indicates  that  the  transmitted  views  can  be  more  efficiently aggregated  as  the  client  requirements  are  more  concentrated in only a few views. This conforms that many applications in which a few major views (i.e., the front sides of objects) are more preferred by users.

In Figs. \ref{3-3} and \ref{3-4}, it is observed that the bandwidth consumption in both the Gaussian and Zipf distributions is also smaller than that in the uniform distribution.

\section{Conclusions}\label{sec:conclusions}

With the recent emergence of 3D-supported TVs, this paper proposes a method for bandwidth-efficient multi-view 3D video multicast over IP networks. By exploiting the DIBR, simulation results show the proposed MMDEA algorithm effectively minimizes total bandwidth consumption by $35\%$ in large networks, and the improvement increases with the number of views and clients, especially in practical scenarios where the clients are more interested in a few select front views in multi-view 3D videos.

\bibliographystyle{IEEEtran}
\bibliography{reference}

\end{document}